\documentclass[11pt,a4paper,oneside]{article}
\usepackage[top=3cm, bottom=3cm, left=2cm, right=2cm]{geometry}
\usepackage[english]{babel}
\usepackage[utf8x]{inputenc}
\usepackage[T1]{fontenc}
\usepackage{enumitem}
\usepackage{amsthm,amsmath,amssymb,mathrsfs,dsfont,mathtools}

\usepackage{bm}
\usepackage[colorinlistoftodos]{todonotes}
\usepackage[colorlinks=true, allcolors=blue]{hyperref}
\usepackage{graphicx}
\usepackage{subcaption}
\usepackage{booktabs}
\usepackage{caption}
\usepackage{cancel}

\usepackage{algpseudocode}
\usepackage{algorithm}

\usepackage[all]{nowidow}

\usepackage{dsfont}

\usepackage{setspace}
\usepackage{tikz} % For the plot
\usetikzlibrary{shapes.geometric, arrows}
\tikzstyle{process} = [rectangle, rounded corners, minimum width=4cm, minimum height=1.2cm, text centered, draw=black]
\tikzstyle{arrow} = [thick,->,>=stealth]

\definecolor{amber}{rgb}{1.0, 0.49, 0.0}

\usepackage[sf]{titlesec}
\usepackage{sectsty}
\allsectionsfont{\centering \normalfont\scshape} % Make all sections centered, the default font and small caps

\usepackage{caption}
\usepackage[round]{natbib}

\usepackage{authblk}
\title{A Bayesian theory for  estimation of biodiversity}
\author[1]{Tommaso Rigon}
\author[2]{Ching-Lung Hsu}
\author[2]{David B. Dunson}
\affil[1]{Department of Economics, Management and Statistics, University of Milano--Bicocca, 20126 Milano, Italy}
\affil[2]{Department of Statistical Science, Duke University, Durham, U.S.A.}

\date{}
\newtheorem{theorem}{Theorem}

\newtheorem{proposition}{Proposition}

\theoremstyle{definition}
\newtheorem{definition}{Definition}
\newtheorem{remark}{Remark}
\newtheorem{example}{Example}

\newcommand \dd  { \,\textup d}   % infintesimal

\begin{document}
\maketitle
\begin{abstract} 
Statistical inference on biodiversity has a rich history going back to RA Fisher. An influential ecological theory suggests the existence of a fundamental biodiversity number, denoted $\alpha$, which coincides with the precision parameter of a Dirichlet process (\textsc{dp}). In this paper, motivated by this theory, we develop Bayesian nonparametric methods for statistical inference on biodiversity, building on the literature on Gibbs-type priors. We argue that $\sigma$-diversity is the most natural extension of the fundamental biodiversity number and discuss strategies for its estimation.  Furthermore, we
develop novel theory and methods starting with an Aldous-Pitman (\textsc{ap}) process, which serves as the building block for any Gibbs-type prior with a square-root growth rate. We propose a modeling framework that accommodates the hierarchical structure of Linnean taxonomy, offering a more refined approach to quantifying biodiversity. The analysis of a large and comprehensive dataset on Amazon tree flora provides a motivating application.
\end{abstract}

\section{Introduction}

The decline in biodiversity represents a significant global concern, with potential implications for entire ecosystems and profound impacts on human well-being \citep[e.g.][]{Ceballos2015}. Consequently, assessing diversity is a primary goal in ecology, although its practical measurement is a notoriously complex task \citep{Colwell2009, Magurran2011}. Taxon richness, the number of taxa within a community, is perhaps the simplest way to describe diversity, although alternative indices such as Simpson and Shannon indices, or Fisher's~$\hat{\alpha}_\textsc{f}$, are often considered. The estimation of richness involves the analysis of \emph{taxon accumulation curves}, a statistical methodology with roots that go back to the seminal work of \citet{Fisher1943}, \citet{Good1953}, and \citet{GoodToulmin1956}. Modern approaches are discussed in \citet{Bunge1993, Colwell2009, Magurran2011}. See also \citet{Zito2023b} for a recent model-based approach to analyzing accumulation curves and estimating richness.

\begin{figure}[tbp]
\centering
\begin{subfigure}{0.4\textwidth}
    \includegraphics[width=\textwidth]{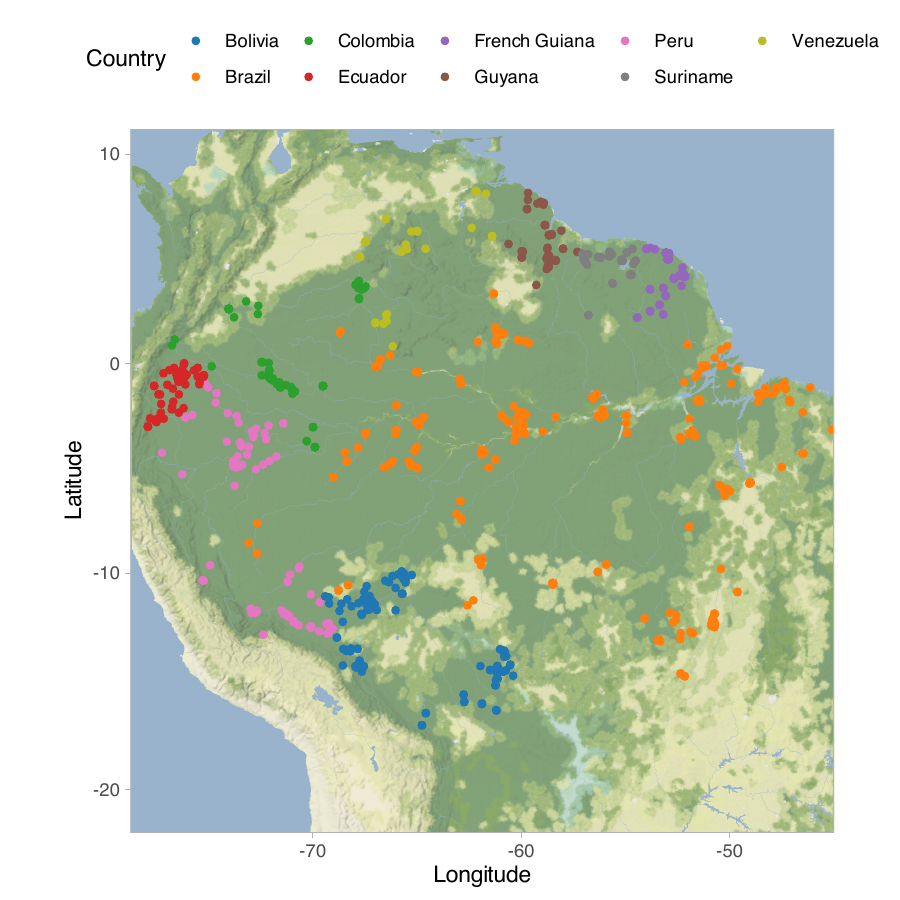}
    \caption{Geographical distribution of the $1170$ tree plots analyzed in \citet{TerSteege2013}. Colors represent the country of the tree plot.}
    \label{fig:sites}
\end{subfigure}
\hfill
\begin{subfigure}{0.55\textwidth}
    \includegraphics[width=\textwidth]{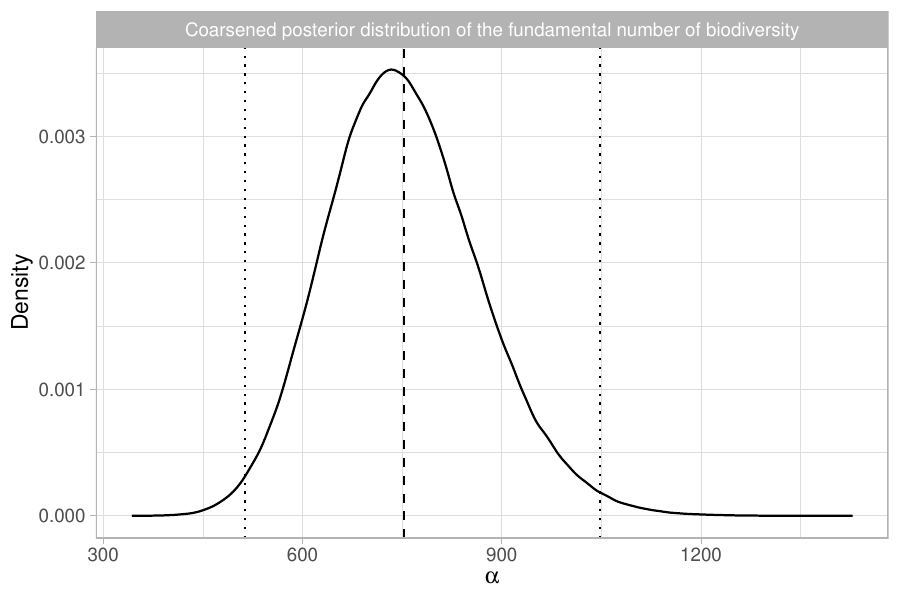}
    \caption{Coarsened posterior distribution ($\rho = 0.01$) of the fundamental biodiversity number $\alpha$, using the Amazonian tree dataset of \citet{TerSteege2013}. The dotted lines represent $98\%$ credible intervals. The dashed line is the posterior mean.}
    \label{fig:post_alpha}
\end{subfigure}
% \caption{Global caption}
\end{figure}

In an influential contribution, \citet{Hubbell2001} postulated the existence of a \emph{fundamental biodiversity number}~$\alpha$, which lies at the core of his unified neutral theory of biodiversity. This number, $\alpha$, represents twice the population size multiplied by the speciation rate.  The theory is conceptually attractive as it encapsulates biodiversity into a single number with a clear biological interpretation. In fact, $\alpha$ is closely linked to accumulation curves, and \citet[][]{Hubbell2001} demonstrated that Fisher's $\hat{\alpha}_\textsc{f}$ is asymptotically equivalent to the fundamental number $\alpha$ for large population size values. The theory has found a successful application in predicting the number of tree species in the Amazon basin, as discussed in \citet{Hubbell2008} and \citet{TerSteege2013}; the data are shown in Figure~\ref{fig:sites}. These studies reported estimates of $\hat{\alpha} = 743$ and $\hat{\alpha} = 754$ for the fundamental biodiversity number, predicting approximately $12,500$ and $15,000$ tree species in the Amazon basin, respectively. 

Interestingly, the fundamental biodiversity number $\alpha$ in Hubbell's theory corresponds to the \emph{precision parameter} of a \textsc{dp} \citep{Hubbell2001}.
The \textsc{dp} has been widely studied and has applications well beyond ecology and biodiversity \citep{Hjort2010, Ghosal2017}. 
Hubbell's theory generated considerable controversy \citep[e.g.][]{McGill2003, Chisholm2004, Ricklefs2006}, being incompatible with data and ignoring mechanisms regarded important by ecologists \citep{McGill2003}. 
In fact, it is well known that \textsc{dp} is
restrictive in depending on a single parameter $\alpha$ and enforcing a logarithmic growth rate for the number of taxa \citep{Lijoi2007, Lijoi2007b}. 
To address these limitations, Gibbs-type priors have emerged as the most natural extension of the \textsc{dp} \citep{Gnedin2005, DeBlasi2015} due to their balance between flexibility and tractability. This class includes
the \textsc{dp}, Pitman-Yor process \citep{Perman1992, Pitman1997}, normalized inverse Gaussian process \citep{Lijoi2005}, and normalized generalized gamma process \citep{Lijoi2007b}. 

We argue that the most natural generalization of Hubbell's fundamental biodiversity number $\alpha$ is the so-called \emph{$\sigma$-diversity} of \citet{Pitman2003}. This retains the main appealing characteristic of the unified neutral theory, corresponding to the ability to describe biodiversity with a single interpretable number, while improving upon Hubbell by allowing for various growth rates for taxon accumulation curves. Classical estimation of $\sigma$-diversity is possible, but, as we shall see, the Gibbs-type framework naturally calls for Bayesian estimates. We contribute to the theory of Gibbs-type priors with novel statistical results and by pointing out connections among the classical work of \citet{Fisher1943}, accumulation curves, and other measures of biodiversity. Moreover, %in Section~\ref{sec:ap}, 
we investigate in depth a Gibbs process we termed Aldous-Pitman after the work of \citet{Aldous1998, Pitman2003}, and we show that a suitable data-augmentation enables posterior inference.

Gibbs-type priors are natural tools for modeling biodiversity when focusing on a single level of the Linnean taxonomy, such as \emph{family, genus} or \emph{species}. However, 
%in modern sampling designs, 
each statistical unit often comprises a collection of $L$ different taxa, which are organized in a nested fashion; see, e.g., \citet{Zito2023c}. As a crude exemplification, one might consider using a different Gibbs-type model for each layer of the Linnean taxonomy. However, this approach would overlook the rich and informative nested structure of the data. Bayesian nonparametric models for such data are less developed. When $L = 2$, a sensible proposal is the enriched Dirichlet process (\textsc{edp}) of \citet{Wade2011}, subsequently extended by \citet{Rigon2025} to the Pitman-Yor case. A more general prior with $L > 2$, and relying on a Pitman-Yor specification, is implicitly employed in \citet{Zito2023c}. We propose a general modelling structure that accounts for the complexities of the available data, focusing on the quantification of biodiversity. We call this novel approach \emph{taxonomic Gibbs-type priors}, which combines the advantages of the enriched Dirichlet process of \citet{Wade2011} with the flexibility of Gibbs processes. In practice, this more refined approach results in taxon-specific indices of biodiversity that might be useful for comparing the biodiversity within branches of the taxonomic tree. 

In Section~\ref{sec:gibbstype}, we discuss Bayesian nonparametric foundations of \citet{Hubbell2001}'s theory of biodiversity. 
In Section~\ref{sec:ap} we use the Aldous-Pitman process to characterize biodiversity. 
In Section~\ref{sec:taxonomic}, we propose taxonomic Gibbs priors.
In Section~\ref{sec:app1}, we 
compare our Bayesian methodology with the analysis of \citet{TerSteege2013}. As depicted in Figure~\ref{fig:post_alpha}, our estimate for $\alpha$ is in close agreement with \citet{Hubbell2008} and \citet{TerSteege2013}, with the benefit of uncertainty quantification. Bayesian inference for the number of tree species is also feasible, as will be discussed in Section~\ref{sec:app1}. In Section~\ref{sec:app2}, we analyze the Amazonian dataset more in-depth through our taxonomic Gibbs-type prior, providing new insights on the within-genera and within-family biodiversity.

\section{Bayesian nonparametric modeling of taxon diversity}\label{sec:gibbstype}
\subsection{Gibbs-type priors}

In this section, we provide an overview of Gibbs-type priors \citep{Gnedin2005}, focusing on their relevance for ecological applications and quantification of biodiversity. For a mathematical exposition, we refer to \citet{DeBlasi2015}, while key
theoretical developments are described in \citet{Lijoi2007, Lijoi2007b,  Lijoi2008b, Lijoi2008, DeBlasi2013}. Our data set comprises taxon labels $X_1,\dots,X_n$, representing, for example, \emph{species} or \emph{families}, and taking values in a set~$\mathds{X}$. We assume that $(X_n)_{n\ge 1}$ is an (ideally) infinite sequence of exchangeable observations which, according to the \emph{de Finetti theorem} \citep{DeFinetti1937},  is equivalent to the following hierarchical specification
\begin{equation}\label{eq:model}
\begin{aligned}
X_n \mid \tilde{p} &\overset{\textup{iid}}{\sim} \tilde{p}, \qquad n \ge 1,\\
\tilde{p} &\sim \mathcal{Q},
\end{aligned}
\end{equation}
where $\tilde{p}$ is a (random) probability measure whose law $\mathcal{Q}$ is the prior distribution in Bayesian statistics. In our setting, we specify a multinomial model for $\tilde{p}$, namely
\begin{equation}\label{eq:multinomial}
\tilde{p}(\cdot) = \sum_{h=1}^{\infty}\pi_h \delta_{Z_h}(\cdot),
\end{equation}
where $\delta_x$ represents the Dirac delta measure at $x$, whereas $\pi_h$ are random probability weights such that $\sum_{h=1}^{\infty}\pi_h = 1$ almost surely, and $(Z_h)_{h \ge 1}$ are random $\mathds{X}$-valued locations, denoting distinct taxon labels. This approach is referred to as \emph{nonparametric} because there are potentially infinitely many probabilities $(\pi_h)_{h \ge 1}$, for which a suitable prior will be specified. We assume that the $Z_h$'s are iid samples from a diffuse probability distribution $P$ on $\mathds{X}$ and that $(Z_h)_{h \ge 1}$ and $(\pi_h)_{h \ge 1}$ are independent. The diffuseness of $P$ implies that the labels $Z_h$'s are all different. Thus, $\tilde{p}$ belongs to the class of proper \emph{species sampling models} \citep[\textsc{ssm}s,][]{Pitman1996}. The baseline measure $P$ serves as a technical tool that simplifies the mathematical exposition, but we are not interested in ``learning'' it. Indeed, without loss of generality, in species sampling problems, we may let $\mathds{X} = \mathds{R}$ and let $P$ be a uniform distribution, meaning that each value $Z_h$ is the \emph{numerical encoding} of the associated taxon. 

\begin{remark} The sampling mechanism in equation~\eqref{eq:model} makes explicit the assumptions behind species sampling models, of which \citet{Hubbell2001}'s theory represents a special case. The core assumptions are: (i) \textsc{ssm}s are tailored for the analysis of individual-based accumulation curves \citep{Gotelli2001} since the taxa $X_n$ are sampled sequentially, one at a time; (ii) taxon labels are independently sampled from $\tilde{p}$ (iid), implying that the probability of observing the $h$th species $Z_h$ remains constant across data points $X_1,\dots, X_n$. These simplifying assumptions facilitate the definition of biodiversity indices, but are often violated in practice, particularly when analyzing meta-community data. For instance, observations $X_1,\dots, X_n$ may originate from different geographical regions, as in \citet{TerSteege2013}, consequently leading to variations in the probability of observing a given taxon. Nevertheless, \textsc{ssm}s serve as a useful approximation of reality, offering robust predictive capabilities, when a reasonable degree of homogeneity across observations is plausible. 
\end{remark}

The primary focus of inference typically lies not in the probabilities $(\pi_h)_{h \ge 1}$ but rather in the combinatorial structure induced by the multinomial model of equation~\eqref{eq:multinomial}. Due to the discrete nature of~$\tilde{p}$, there will be identical values among $X_1,\dots,X_n$ with positive probability, comprising a total of $K_n = k$ distinct values with labels $X_1^{*},\dots,X_k^{*}$ and frequencies $n_{1},\dots,n_{k}$ such that $\sum_{j=1}^k n_j=n$. The random variable $K_n$ represents the observed taxon richness, whereas $n_1,\dots,n_k$ denote the associated abundances. The ties among the observations $X_1\dots,X_n$ induce a random partition $\Psi_n = \{C_1,\dots,C_k\}$ of the indices $\{1,\dots,n\}$, where $C_j=\{i: X_i = X_j^{*}\}$ and $n_j = |C_j|$. A species sampling model $\tilde{p}$ is a \emph{Gibbs-type process} if the law of the random partition, called \emph{Gibbs partition}, is such that
\begin{equation}
\label{eqn:Gibbs_type}
\Pi_n\left(n_{1},\dots,n_{k}\right)=\mathds{P}\left(\Psi_n = \{C_1,\dots,C_k\}\right) = V_{n,k}\prod_{j=1}^{k}(1-\sigma)_{n_j-1},
\end{equation}
where $(a)_{n}= a (a + 1)\cdots (a + n - 1)$ denotes a rising factorial, $\sigma < 1$ is a discount parameter, and $V_{n,k}$ are non-negative weights satisfying the forward recursion $V_{n,k}=\left(n-\sigma k\right)V_{n+1,k} + V_{n+1,k+1}$ for any $n\geq 1$ and $1\leq k\leq n$. $\Pi_n\left(n_{1},\ldots,n_{k}\right)$ is called the \emph{exchangeable partition probability function} (\textsc{eppf}) and, in conjunction with $P$, characterizes the distribution of random probability measure $\tilde{p}$ \citep{Pitman1996}. Of considerable  importance is the predictive distribution induced by $\Pi_n$ \citep{Lijoi2007}: 
\begin{equation}\label{eq:predictive}
\mathds{P}(X_{n+1} \in \cdot \mid X_1,\dots,X_n) = \frac{V_{n+1, k+1}}{V_{n,k}}P(\cdot) + \frac{V_{n+1, k}}{V_{n,k}}\sum_{j=1}^k(n_j - \sigma)\delta_{X^*_j}(\cdot).
\end{equation}
The above predictive rule provides a sampling mechanism for the data $X_1,\dots,X_n$. The term $\mathds{P}(X_{n+1} = \text{``new''} \mid X_1,\dots,X_n) = V_{n+1, k+1} / V_{n,k}$ denotes the probability of discovering a new taxon, i.e., the probability of $X_{n+1}$ not being observed among the previous $X_1,\dots,X_n$. Conversely, the probability $\mathds{P}(X_{n+1} = \text{``old''} \mid X_1,\dots,X_n) = 1 - V_{n+1, k+1} / V_{n,k}$ corresponds to a Bayesian estimate of the \emph{sample coverage}, namely the fraction of already observed taxa. The probability of re-observing the $j$th taxon $X^*_j$ is proportional to $n_j - \sigma$, explaining why $\sigma < 1$ is sometimes referred to as the ``discount parameter'', as it diminishes (or augments) the observed frequencies. Notable Gibbs-type priors are discussed below.

\begin{example}[Dirichlet--multinomial, case $\sigma < 0$]
	\label{ex_dm} Consider $\sigma < 0$, and let $H \in \mathds{N}$. A valid set of Gibbs coefficients is defined as:
	\begin{equation}
	\label{eq:dm}
	V_{n, k}(\sigma, H) := \frac{|\sigma|^{k-1}\prod_{j=1}^{k-1}(H - j)}{(H|\sigma| +1)_{n-1}} 
	\mathds{1}(k \le H),
	\end{equation}
	where $\mathds{1}(\cdot)$ denotes the indicator function. This model assumes a finite number of taxa at a population level because the in-sample species richness $K_n = k$ is upper-bounded by $H$. Its predictive scheme is:
\begin{equation*}
	\mathds{P}(X_{n+1} \in \cdot \mid H, X_1,\dots,X_n) = \frac{(H-k)|\sigma|}{H|\sigma| + n}P(\cdot) + \frac{1}{H|\sigma| + n} \sum_{j=1}^k (n_j + |\sigma|)\delta_{X^*_j}(\cdot).
\end{equation*}
Thus, the probability of discovering a new taxon is zero whenever $H = k$. The corresponding process is called Dirichlet-multinomial, denoted by $\tilde{p} = \sum_{h=1}^H W_h \delta_{Z_h}$, with $(W_1,\dots,W_{H-1}) \sim \text{Dir}(|\sigma|,\dots,|\sigma|)$. 
\end{example}

\begin{example}[Dirichlet process, case $\sigma = 0$] 
	\label{ex_dp}
For $\sigma = 0$ and $\alpha > 0$, a valid set of Gibbs coefficients is:
	\begin{equation}
	\label{eq:dp}
	V_{n, k}(\alpha) := \frac{\alpha^k}{(\alpha)_n}.
	\end{equation}
	The corresponding process $\tilde{p}$ is the Dirichlet process of \citet{Ferguson1973}, with precision parameter $\alpha$, the fundamental biodiversity parameter of \citet{Hubbell2001}. The general predictive rule in~\eqref{eq:predictive} becomes:	\begin{equation*}
	\mathds{P}(X_{n+1} \in \cdot \mid \alpha, X_1,\dots,X_n) = \frac{\alpha}{\alpha + n}P(\cdot) + \frac{1}{\alpha + n} \sum_{j=1}^k n_j\delta_{X^*_j}(\cdot),
	\end{equation*}
	the \citet{Blackwell1973} urn scheme. The Dirichlet process $\tilde{p}$ admits a \emph{stick-breaking} representation \citep{Sethuraman1994}: $\tilde{p}= \sum_{h=1}^\infty \xi_h \delta_{Z_h}$ with $\pi_h = \xi_h \prod_{\ell < h}(1 - \xi_\ell)$ and $\xi_h \overset{\text{iid}}{\sim}\text{Beta}(1, \alpha)$ for $h \ge 1$, with $\pi_1 = \xi_1$. This highlights that (i) the \textsc{dp} assumes infinitely many taxa at a population level, and (ii) the probabilities of these taxa decrease exponentially fast at a rate determined by $\alpha$. 
\end{example}

\begin{example}[$\sigma$-stable Poisson-Kingman process, case $\sigma \in (0,1)$] \label{ex:pk}
For $\sigma \in (0,1)$ and $\gamma > 0$, we have 
%a valid set of Gibbs coefficients is defined as
	\begin{equation}
	\label{eq:pk}
	V_{n, k}(\sigma, \gamma) := \frac{\sigma^k\gamma^k}{\Gamma(n-k\sigma)f_{\sigma}\big(\gamma^{-1/\sigma}\big)}\int_{0}^{1}s^{n-1-k\sigma}f_{\sigma}\big(\left(1-s\right)\gamma^{-1/\sigma}\big)\dd s,
	\end{equation}
where $\Gamma(\cdot)$ is the gamma function and $f_\sigma(t) = (\pi)^{-1} \sum_{h=1}^\infty(-1)^{h+1}\sin(h \pi \sigma )\Gamma(h\sigma + 1) / t^{h\sigma + 1}$ represents the density of a $\sigma$-stable distribution. The corresponding process $\tilde{p}$ is a $\sigma$-stable Poisson-Kingman (\textsc{pk}) process, conditional on ``total mass'' $\gamma^{-1/\sigma}$ \citep{Kingman1975, Pitman2003}. The distribution of the associated probability weights $(\pi_h)_{h \ge 1}$ is complex, but implies infinitely many species.
\end{example}

Remarkably, the aforementioned set of weights $V_{n, k}(\sigma, H)$, $V_{n, k}(\alpha)$ and $V_{n, k}(\sigma, \gamma)$ form the foundation of any Gibbs-type prior. In fact, Gibbs partitions can always be expressed as a mixture with respect to the parameters $H$, $\alpha$, and $\gamma$. This result was established by \citet{Gnedin2005}.
 
\begin{proposition}[\citet{Gnedin2005}]
\label{gibbs_characterization}
 The Gibbs coefficients $V_{n,k}$ satisfy the recursive equation $V_{n,k}=\left(n-\sigma k\right)V_{n+1,k} + V_{n+1,k+1}$ for any $n\geq 1$ and $1\leq k\leq n$ in the following three cases:
\begin{enumerate}
\item If $\sigma < 0$, whenever $V_{n,k} = \sum_{h=1}^\infty V_{n, k}(\sigma, h) p(h)$,
for some discrete random variable $H \in \mathds{N}$ with probability distribution function $p(h)$, where the $V_{n, k}(\sigma, h)$'s are defined as in equation~\eqref{eq:dm}. 
\item If $\sigma = 0$, whenever $V_{n,k} = \int_{\mathds{R}^+} V_{n, k}(\alpha) p(\mathrm{d}\alpha)$, for some positive random variable $\alpha$ with probability measure $p(\mathrm{d}\alpha)$, where the $V_{n, k}(\alpha)$'s are defined as in equation~\eqref{eq:dp}.
\item If $\sigma \in (0,1)$, whenever $V_{n,k} = \int_{\mathds{R}^+} V_{n, k}(\sigma, \gamma) p(\mathrm{d}\gamma)$, for some positive random variable $\gamma$ with probability measure $p(\mathrm{d}\gamma)$, where the $V_{n, k}(\sigma, \gamma)$'s are defined as in equation~\eqref{eq:pk}. 
\end{enumerate}
\end{proposition}

Any Gibbs-type process can be represented hierarchically, involving a suitable prior distribution for the key parameters $H$, $\alpha$, and $\gamma$. Prior distributions for $H$ in the $\sigma < 0$ case are discussed in \citet{Gnedin2010} and \citet{DeBlasi2013}; see also \citet{Miller2018} for applications to mixture models and \citet{Legramanti2022} for employment in stochastic block models. In the $\sigma = 0$ case, a popular choice is the semi-conjugate Gamma prior for $\alpha$ as in \citet{Escobar1995}, with 
the Stirling-gamma prior of \citet{Zito2024} a recent alternative. Finally, in the $\sigma \in (0, 1)$ regime, a polynomially tilted stable distribution prior for $\gamma^{-1/\sigma}$ leads to the Pitman-Yor process \citep{Pitman1997}, whereas an exponentially tilted stable density leads to the normalized generalized gamma process \citep{Lijoi2007b}; refer to \citet{Lijoi2008, Favaro2009} for a thorough investigation and to \citet{Favaro2015} for a general class of priors for $\gamma$.

\subsection{The quantification of biodiversity}

The simplest measure of biodiversity is arguably the taxon \emph{richness}. In our notation, we observe $K_n = k$ different taxa among $n$ data points $X_1, \dots, X_n$; we refer to this value as the in-sample richness. A priori, the distribution of $K_n$ induced by a Gibbs-type prior has a simple form 
\begin{equation*}
\mathds{P}(K_{n}=k)=V_{n,k}\frac{\mathscr{C}(n,k;\sigma)}{\sigma^k},
\end{equation*}
where $\mathscr{C}(n,k;\sigma)$ denotes a generalized factorial coefficient \citep{Charalambides2002}; the case $\sigma = 0$ is recovered as the limit of the above formula, since $\lim_{\sigma \rightarrow 0}  \mathscr{C}(n,k;\sigma)\sigma^{-k} = |s(n, k)|$, which is the signless Stirling number of the first kind. 
The expected values $\mathds{E}(K_1),\dots,\mathds{E}(K_n)$ can be interpreted as a model-based \emph{rarefaction} curve \citep{Zito2023b}. Moreover, we may want to predict the number of new taxa $K_{n+m} \ge k$ within a future sample $X_{n+1},\dots, X_{n+m}$; we call this value the out-of-sample richness. The posterior distribution of $K_{n+m}$ given the data $K_n = k$, or equivalently of the number of previously unobserved taxa $K_m^{(n)} = K_{n+m} - K_n$, has been derived by \citet{Lijoi2007} and is 
\begin{equation*}
\mathds{P}(K_m^{(n)}= j \mid X_1,\dots,X_n)=\frac{V_{n + m, k + j}}{V_{n,k}}\frac{\mathscr{C}(m, j;\sigma, -n + k\sigma)}{\sigma^j}, \qquad j=0,\dots,m,
\end{equation*}
where the term $\mathscr{C}(m, j;\sigma, -n + k\sigma)$ is the noncentral generalized factorial coefficient; see \citet{Lijoi2007}. %The result holds for $\sigma < 1$, thus including negative values, as remarked in \citet{Lijoi2008b}. 
The collection $\mathds{E}(K_{n+1} \mid K_n = k), \dots, \mathds{E}(K_{n+m} \mid K_n = k)$ represents a model-based \emph{extrapolation} of the accumulation curve. Note that $K_n = k$ is a sufficient statistic for predictions. 

\begin{example}[Dirichlet-multinomial, cont'd]\label{ex:dm} For the Dirichlet-multinomial, the rarefaction curve is 
\begin{equation*}
\mathds{E}(K_n \mid H) = H - H \frac{(H|\sigma| - |\sigma|)_n}{(H|\sigma|)_{n}},
\end{equation*}
as shown in \citet[Chap. 3]{Pitman2006}. Moreover, in the Appendix we provide a simple proof to show that:
\begin{equation*}
\mathds{E}(K_{n+m} \mid H, X_1,\dots,X_n) = H - (H-k)\frac{(n + H|\sigma| - |\sigma|)_m}{(n + H|\sigma|)_m}.
\end{equation*}

\end{example}

\begin{example}[Dirichlet process, cont'd]

In the Dirichlet process case, the rarefaction and the extrapolation curves are given by:
\begin{equation*}
\mathds{E}(K_n \mid \alpha) = \sum_{i=1}^n\frac{\alpha}{\alpha + i - 1}, \qquad \mathds{E}(K_{n + m}\mid \alpha, X_1,\dots,X_n) = k + \sum_{i=1}^m \frac{\alpha}{\alpha + n + i -1}.
\end{equation*}
For their practical implementation, one can use $\sum_{i=1}^n 1/ (a + i - 1) = \psi(a +n) - \psi(a)$ for any $a >0$,  where $\psi(\cdot)$ is the digamma function. See, for example, \citet{Zito2023b}.
\end{example}

While rarefaction and extrapolation curves are valuable tools \citep{Gotelli2001}, it is useful to summarize biodiversity with a single number. The concept of \emph{richness}, defined as $\lim_{n \rightarrow \infty} K_n$, 
is problematic since $\lim_{n\to \infty} K_n = \infty$ for 
 $\sigma \ge 0$ regardless of the observed data, whereas $K_n$ remains finite for $\sigma < 0$. Avoiding $\sigma \ge 0$ is tempting, but leads to poor fit and predictions for some datasets. Gibbs-type priors with positive $\sigma \ge 0$ often excel in predicting future values $K_m^{(n)}$ for highly diverse taxa, compared to models with $\sigma \le 0$ \citep{Lijoi2007, Favaro2009}. Additionally, the total number of taxa present in a given area can be estimated even in models where $\sigma \ge 0$, e.g. following the strategy of \citet{TerSteege2013}. This discussion motivates embracing an
 alternative concept of diversity, called $\sigma$-diversity, introduced by \citet{Pitman2003}, which encompasses \citet{Hubbell2001}'s fundamental biodiversity number when $\sigma = 0$.
\begin{proposition}[$\sigma$-diversity, \citet{Pitman2003}]
\label{diversity}
 Let $K_n$ be the number of distinct values arising from a Gibbs-type prior in~\eqref{eqn:Gibbs_type}:
\begin{enumerate}
\item Let $\sigma < 0$ and $V_{n, k}(\sigma, H)$ be defined as in equation~\eqref{eq:dm}, then $K_n \rightarrow H$ almost surely;
\item Let $\sigma = 0$ and $V_{n, k}(\alpha)$ be defined as in equation~\eqref{eq:dp}, then $K_n / \log(n) \rightarrow \alpha$ almost surely;
\item Let $\sigma \in (0,1)$ and $V_{n, k}(\sigma, \gamma)$ be defined as in equation~\eqref{eq:pk}, then $K_n / n^\sigma \rightarrow \gamma$ almost surely.
\end{enumerate}
For a generic set of weights $V_{n,k}$, let $c_\sigma(n)$ be a function such that $c_\sigma(n) = 1$ if $\sigma <0$, $c_\sigma(n) =\log(n)$ if $\sigma = 0$, and $c_\sigma(n) = n^{\sigma}$ if $\sigma\in\left(0,1\right)$. Then as $n \rightarrow \infty$
\begin{equation}
\label{eqn:asymptotic_diversity}
\frac{K_{n}}{c_\sigma(n)}\overset{a.s.}{\longrightarrow} S_{\sigma}.
\end{equation}
The random variable $S_{\sigma}$ is called \emph{$\sigma$-diversity} and its distribution coincides with the prior for $H$, $\alpha$ and~$\gamma$, respectively, implied by the mixture representation of the weights $V_{n,k}$ in Proposition~\ref{gibbs_characterization}.
\end{proposition}

The $\sigma$-diversity can be seen as a richness measure that has been appropriately rescaled. Proposition~\ref{diversity} highlights the central role of the Dirichlet-multinomial, Dirichlet, and $\sigma$-stable \textsc{pk} processes among Gibbs-type priors. It shows that their parameters $H$, $\alpha$, and $\gamma$ are biodiversity indices. In these three processes, $\sigma$ diversity is deterministic and assumed to be known. However, $\sigma$-diversities $H$, $\alpha$, or $\gamma$ are typically unknown, and can be estimated employing a prior distribution, leading to a Gibbs-type process for $\tilde{p}$ thanks to Proposition~\ref{gibbs_characterization}. In light of this, the posterior law of $\sigma$-diversity is a key quantity for measuring biodiversity. In addition, the posterior distribution of $S_\sigma$ has an elegant connection with accumulation curves, as shown in the following theorem.
\begin{theorem}
\label{posterior_diversity}
Let $X_1,\dots,X_{n+m}$ be a sample from a Gibbs-type prior~\eqref{eqn:Gibbs_type} with $K_{n+m}$ distinct values and let the function $c_\sigma(m)$ be defined as in Proposition~\ref{diversity}. Then as $m \rightarrow \infty$
\begin{equation}
\left(\frac{K_{n + m}}{c_\sigma(m)} \mid X_1,\dots,X_n \right) \overset{a.s.}{\longrightarrow} S_{\sigma}, 
\end{equation}
The random variable $S_{\sigma}$ is the \emph{$\sigma$-diversity} and its distribution coincides with the posterior for $H$, $\alpha$ and~$\gamma$, respectively. Moreover, $K_n = k$ is a sufficient statistic for $S_\sigma$ given the data $X_1,\dots,X_n$.
\end{theorem}

 Theorem~\ref{posterior_diversity} states that the posterior law of $S_\sigma$ coincides with the $\sigma$-diversity associated with extrapolation of the accumulation curve. Related results for the normalized generalized gamma model are in \citet{Favaro2012}. In practice, deriving the posterior distribution of $\sigma$-diversities $H$, $\alpha$ and $\gamma$ is based on the Bayes theorem, where the \textsc{eppf} in \eqref{eqn:Gibbs_type} acts as \emph{likelihood function}. %For example, suppose $\sigma = 0$ and , then the posterior law for the $\sigma$-diversity is $p(\mathrm{d}\alpha \mid X_1,\dots,X_n) \propto p(\mathrm{d}\alpha) V_{n, k}(\alpha)$. 
Notably, in Gibbs-type priors, the abundances $n_1,\dots,n_k$ appearing in~\eqref{eqn:Gibbs_type} do not provide insights about the diversity, and they do not appear in the posterior for $S_\sigma$ because the number of observed species $k$ is a sufficient statistic. This would not be the case for general species sampling models beyond the Gibbs-type. 

\begin{example}[Dirichlet-multinomial, cont'd] Bayesian inference for the richness $H$, i.e. the $\sigma < 0$ case, is based on the posterior distribution
\begin{equation*}
p(h \mid X_1,\dots,X_n) \propto p(h)\frac{|\sigma|^{k-1}\prod_{j=1}^{k-1}(h - j)}{(h|\sigma| +1)_{n-1}}, \qquad h=k, k+1,\dots,
\end{equation*}
where $p(h)$ denotes a discrete prior distribution for $H \in \{1,2,\dots\}$. If the prior has bounded support, sampling from the posterior is trivial. In general, acceptance-rejection or truncation strategies can be considered. When $\sigma = -1$, shifted geometric and shifted Poisson prior specifications have been investigated in \citet{DeBlasi2013}, whereas \citet{Gnedin2010} proposed a heavy-tailed prior distribution, which leads to a closed-form expression for the posterior distribution of $H$ and the weights $V_{n,k}$.
\end{example}

\begin{example}[Dirichlet-process, cont'd]
Bayesian inference about the fundamental biodiversity number $\alpha$, i.e. the $\sigma = 0$ case, is based on the posterior distribution
\begin{equation*}
p(\mathrm{d}\alpha \mid X_1,\dots,X_n) \propto p(\mathrm{d}\alpha) \frac{\alpha^k}{(\alpha)_n},
\end{equation*}
where $p(\mathrm{d}\alpha)$ denotes the prior distribution for $\alpha$. As noted in \citet{Zito2023b}, this problem is equivalent to a logistic regression with an offset. The choice $\alpha \sim \text{Gamma}(a,b)$ has been advocated by \citet{Escobar1995} because it leads to a semi-conjugate Gibbs-sampling scheme. More recently, \citet{Zito2024} proposed the (asymptotically equivalent) Stirling-gamma prior, which has the advantage of being highly interpretable, especially within the context of species sampling models. 
\end{example}

\begin{remark}\label{rem:diversities}
The discount parameter $\sigma$ characterizes the asymptotic behavior of the Gibbs partition. Although one could theoretically estimate $\sigma$ from data using a prior distribution, leading to a species sampling model beyond the Gibbs type, we argue that the choice of $\sigma$ should be regarded as a model selection problem. This avoids difficulties
in interpretation, since comparing $\sigma$-diversities across locations makes sense only if they are based on the same asymptotic regime. In practice, based on the data, $\sigma$ can be chosen from the following values: $\sigma = -1$ (Dirichlet distribution with uniform weights), $\sigma = 0$ (Dirichlet process) and $\sigma = 1/2$ (Aldous-Pitman process).
\end{remark}

\begin{remark}\label{rem:fisher}
Fisher's~$\hat{\alpha}_\textsc{f}$ \citep{Fisher1943} is an estimator for the parameter $\alpha$ of the Dirichlet process \citep{Hubbell2001}. We compare Fisher's $\hat{\alpha}_\textsc{f}$ with the maximum likelihood estimator $\hat{\alpha}_\textsc{ml} = \arg\max_{\alpha} \Pi_n(n_1,\dots,n_k \mid \alpha) = \arg\max_{\alpha} \alpha^k / (\alpha)_n$ of the diversity, with these estimators solving 
\begin{equation*}
\hat{\alpha}_\textsc{f}\log\left(1 + \frac{n}{\hat{\alpha}_\textsc{f}}\right) = k \qquad \text{ and } \qquad \sum_{i=1}^n \frac{\hat{\alpha}_\textsc{ml}}{\hat{\alpha}_\textsc{ml} + i - 1} = k.
\end{equation*}
For large $n$, the two estimates are nearly identical, due to the inequalities $\alpha\log\left(1 + n / \alpha \right) \le \sum_{i=1}^n \alpha/ (\alpha + i - 1) \le 1 + \alpha\log\left(1 + n / \alpha \right)$ \citep[see, e.g.,][Proposition 4.8]{Ghosal2017}. This striking similarity between $\hat{\alpha}_\textsc{f}$ and $\hat{\alpha}_\textsc{ml}$ results because the \textsc{eppf} of a Dirichlet process can be regarded as a conditional likelihood for Fisher's model, in which the sample size $n$ is fixed and not random, contrary to Fisher's original formulation; see \citet{McCullagh2016} for a detailed historical account and further considerations. Hence, the posterior law of $\alpha$ provides a Bayesian Fisher's $\hat{\alpha}_\textsc{f}$. The Bayesian perspective not only establishes an insightful link between Fisher's $\hat{\alpha}_\textsc{f}$ and accumulation curves through Proposition~\ref{diversity} and Theorem~\ref{posterior_diversity}, but also facilitates uncertainty quantification and testing.
\end{remark}

\subsection{Relationship between $\sigma$-diversity and other biodiversity measures}\label{sec:diversities}

We show that many classical biodiversity indices can be expressed in terms of $H$, $\alpha$, or $\gamma$, based on the connection with the accumulation curves discussed earlier. Consider the popular
Simpson similarity index $\mathcal{S}$ \citep[e.g.,][]{Colwell2009}, which is the probability that two randomly chosen individuals belong to the same taxa. Within the framework of species sampling models, 
%\begin{equation*}
$\mathcal{S} := \sum_{h=1}^\infty \pi_h^2,$
and due to exchangeability 
%\qquad \text{and so} \qquad 
$\mathds{E}(\mathcal{S}) = \mathds{P}(X_1 = X_2).$
%\end{equation*}
In Gibbs-type priors, it follows that $\mathds{E}(\mathcal{S}) = V_{2, 1}$. We can specialize this formula in the Dirichlet multinomial and Dirichlet process case, yielding respectively
\begin{equation*}
\mathds{E}\{\mathcal{S}(\sigma, H)\mid H\} = V_{2, 1}(\sigma, H) = \frac{1}{1 + H |\sigma|}, \qquad \mathds{E}\{\mathcal{S}(\alpha) \mid \alpha\} = V_{2, 1}(\alpha) = \frac{1}{1 + \alpha}.
\end{equation*}
Thus, there is an inverse relationship between the expected Simpson similarity and the $\sigma$-diversity. The following section shows a similar relationship in the polynomial regime, when $\sigma = 1/2$. More generally, most biodiversity indices can be expressed as 
%\begin{equation*}
$\mathcal{I}_g := \sum_{h=1}^\infty g(\pi_h),$
%\end{equation*}
for some function $g :(0,1)\rightarrow \mathds{R}^{+}$. For instance, $g(\pi) = \pi^2$ corresponds to the Simpson index, $g(\pi) = -\pi\log{\pi}$ is the Shannon diversity, and $g(\pi) = \pi^\tau$ for some $\tau > 0$ is the (unnormalized) Tsallis or Hill diversity \citep{Colwell2009, Magurran2011}. The expectations $\mathds{E}(\mathcal{I}_g)$ are a function of $H, \alpha$ and $\gamma$. Additionally, their relationship with $\mathcal{I}_g$ can often be explicitly determined: if the random probabilities $(\pi_h)_{h \ge 1}$ are in \emph{size-biased order}, then we have the simplification
$\mathds{E}(\mathcal{I}_g) = \mathds{E}\left\{\sum_{h=1}^\infty g(\pi_h)\right\} = \mathds{E}\{g(\pi_1)/ \pi_1\}$. For example, in the \textsc{dp} case, the stick-breaking weights are in size-biased order, and we have $\pi_1 \sim \text{Beta}(1, \alpha)$, which makes computations straightforward. 
%We refer, e.g., to \citet{Pitman2003} for a more detailed discussion. 

\subsection{Model validation}\label{sec:check}

We propose two approaches to assess the fit of the model. The first approach compares the observed distinct values $K_1,\dots,K_n$ with the model-based rarefaction curve $\mathds{E}(K_1),\dots,\mathds{E}(K_n)$. As shown above, expectations $\mathds{E}(K_i)$ can often be computed explicitly, and are typically evaluated given an estimate for the diversity parameter, e.g. $\mathds{E}(K_i \mid \hat{\alpha}_\textsc{ml})$ in the $\sigma = 0$ case. Due to exchangeability, expectations $\mathds{E}(K_i)$ do not depend on the order of the data. However, in most cases, the data $X_1,\dots,X_n$ are not observed in a specific order. Thus, a common practice is to reshuffle the order of $X_1,\dots,X_n$ and then consider the averages $\bar{K}_1,\dots,\bar{K}_n$, over all possible permutations of the data. Conveniently, for this combinatorial problem, there exists an explicit solution, which is:
\begin{equation*}
\bar{K}_i = k - \binom{n}{i}^{-1}\sum_{j=1}^k\binom{n - n_j}{i}, \qquad i=1,\dots,n,
\end{equation*}
where $K_n = k$ and $n_1,\dots,n_k$ are the abundances; see \citet{Smith1977, Colwell2012}. The values $\bar{K}_i$ define the ``classical rarefaction'', which can be regarded as a frequentist nonparametric estimator arising from a multinomial model. Our Bayesian nonparametric approach instead depends on a few parameters, imposing some rigidity on the functional shape of the rarefaction, which is critical for extrapolation. Graphically comparing $\bar{K}_1,\dots,\bar{K}_n$ with $\mathds{E}(K_1),\dots,\mathds{E}(K_n)$ can give a sense of the suitability of the chosen model. We show a concrete example in Section~\ref{sec:app1}. 

The second model-checking approach is even simpler and has been used, e.g. in \citet{Favaro2021}; see also \citet{Thisted1987} for early ideas. The abundances $n_1,\dots,n_k$ can be reformulated in terms of frequency counts $m_1,\dots,m_n$, where $m_r$ is the number of taxa appearing with frequency $r$ in the sample. Hence, $m_1$ represents the number of singletons, $m_2$ is the number of doubletons, etc. We denote by $M_{1,n},\dots,M_{n,n}$ the associated random variables. To assess goodness of fit, we compare empirical counts $m_1,\dots,m_n$ with their model-based expectations $\mathds{E}(M_{1,n}),\dots,\mathds{E}(M_{n,n})$. 
In the Dirichlet process case, these expectations have a simple analytical formula
\begin{equation*}
\mathds{E}(M_{r,n} \mid \alpha) = \alpha \frac{(\alpha)_{n-r}}{(\alpha)_{n}}\binom{n}{r}(r -1)!, \qquad r=1,\dots,n,
\end{equation*}
and therefore when $r=1$, corresponding to the expected singletons implied by \citet{Hubbell2001} theory, we get $\mathds{E}(M_{1,n} \mid \alpha) = n \alpha / (n + \alpha) = n \: \text{pr}(X_{n+1} = \text{``new''} \mid \alpha, X_1,\dots,X_n)$. 
The general analytical formulas for the expectations $\mathds{E}(M_{r,n})$ for any prior of Gibbs type are given in \citet{Favaro2013}. These expectations can be approximated via Monte Carlo, by sampling $X_1,\dots,X_n$ from the urn scheme in equation~\eqref{eq:predictive}.

\section{The Aldous-Pitman process ($\sigma = 1/2$)}\label{sec:ap}

The Dirichlet multinomial and Dirichlet process cases, corresponding to $\sigma \le 0$, have been extensively investigated in Bayesian nonparametrics. There exist suitable priors and well-established estimation procedures for $H$ and $\alpha$, as discussed in Section~\ref{sec:gibbstype}. In contrast, relatively less attention has been devoted to the case $\sigma > 0$, which corresponds to rapidly growing accumulation curves, with notable exceptions such as \citet{Lijoi2007b} and \citet{Favaro2009}. The main mathematical challenge lies in the density $f_\sigma(\cdot)$ in equation~\eqref{eq:pk}, which lacks a simple analytical expression. However, in the $\sigma = 1/2$ case, this density becomes $f_{1/2}(t) = (\sqrt{4\pi})^{-1} t^{-3/2} \exp\{-1/(4t)\}$, significantly simplifying the calculations. This leads to the definition of what we term the Aldous-Pitman process.

\begin{definition}[Aldous-Pitman process] Let $\gamma > 0$, and $P$ be a diffuse probability measure on $\mathds{X}$. Additionally, let $(Z_n)_{n\ge1}$ be iid samples from $P$ and $(Y_n)_{n \ge 1}$ and be iid samples from a standard Gaussian distribution. Define a species sampling model $\tilde{p} = \sum_{h=1}^{\infty}\pi_h \delta_{Z_h}$ where
\begin{equation*}
\pi_h = R_{h-1} - R_h, \qquad R_h =  \frac{\gamma^2/2}{\gamma^2/2 + \sum_{j=1}^h Y_j^2}, \qquad h \ge 1,
\end{equation*}
and $R_0 = 1$. We will say that $\tilde{p} \mid \gamma$ follows an Aldous-Pitman process with parameters $\gamma$ and $P$.%, denoted as $\tilde{p} \mid \gamma \sim \textsc{ap}(\gamma, P)$.
\end{definition}

The Aldous-Pitman process, introduced by \citet{Aldous1998}, is a special case of the $\sigma$-stable \textsc{pk} process described in Example~\ref{ex:pk}, when $\sigma = 1/2$  \citep{Pitman2003}. Consequently, the Aldous-Pitman process results in a Gibbs partition, where the parameter $\gamma$ corresponds to the $\sigma$-diversity. Notably, the weights satisfy $\pi_h > \pi_{h-1}$, where $R_h = \sum_{\ell > h} \pi_\ell$ represents the cumulative probability of rare taxa, which is directly influenced by the $\sigma$-diversity parameter~$\gamma$. Furthermore, the associated Gibbs-type coefficients $V_{n,k}$ can be explicitly obtained.

\begin{example}[Aldous-Pitman process, case $\sigma = 1/2$] \label{ex:ap}
Suppose $\sigma =1/2$ and let $\gamma > 0$. Then a valid set of Gibbs coefficients is given by
	\begin{equation}
	\label{eq:ap}
	V_{n, k}(\gamma) = 2^{n - k/2 - 1/2} \gamma^{k-1}h_{k+1-2n}(\sqrt{2}\gamma),
	\end{equation}
where $h_\nu(t) = \{2\Gamma(-\nu)\}^{-1} \sum_{h=0}^\infty(-t)^h/h!\Gamma\{(h-\nu)/2\}$ denotes the \emph{Hermite function} of order $\nu \in \mathds{R}$ \citep[][\S 10.2]{Lebedev1965}.  The corresponding process $\tilde{p}$ is an Aldous-Pitman process, and the $\sigma$-diversity is $\gamma$. Moreover, the associated urn-scheme is as follows:
\begin{equation*}
\mathds{P}(X_{n+1} \in \cdot \mid \gamma, X_1,\dots,X_n) = \sqrt{2}\gamma \frac{h_{k-2n}(\sqrt{2}\gamma)}{h_{k+1-2n}(\sqrt{2}\gamma)}P(\cdot) + 2\frac{h_{k-2n -1}(\sqrt{2}\gamma)}{h_{k+1-2n}(\sqrt{2}\gamma)}\sum_{j=1}^k(n_j - 1/2)\delta_{X^*_j}(\cdot).
\end{equation*}
It can be shown \citep{Pitman2003} that the expected Simpson index of an Aldous-Pitman process is
\begin{equation*}
\mathds{E}(\mathcal{S}(\gamma) \mid \gamma) = \mathds{P}(X_1 = X_2 \mid \gamma) = \mathds{E}\big(e^{-\gamma\sqrt{V}}\big), \qquad V \sim \text{Exp}(1),
\end{equation*}
revealing the close relationship between the $\sigma$-diversity $\gamma$ and the Simpson index.
\end{example}

The \textsc{ap} process serves as a building block for Gibbs-type priors with growth rate $\sqrt{n}$, having a $\sigma$-diversity $\gamma$. So far, two specific priors for $\gamma$ have been investigated: the prior implicitly utilized in the Pitman--Yor process \citep{Favaro2009} and the one implied by the normalized generalized gamma (\textsc{ngg}) process \citep{Lijoi2007b}. Both options yield tractable Gibbs coefficients $V_{n,k}$ for all values of $\sigma \in (0,1)$. Furthermore, when $\sigma = 1/2$ these priors reduce to the following specifications
\begin{equation}
\begin{aligned}
\label{eq:prior_ap}
\gamma^2 &\sim \text{Gamma}(\theta + 1/2, 1/4), \qquad &&\text{(Pitman-Yor)} \\
\gamma^{-2} &\sim \text{Inverse-Gaussian}(1/\sqrt{2}, \beta / \sqrt{2}), \qquad &&\text{(\text{Normalized inverse Gaussian})} \\
\end{aligned}
\end{equation}
where $\theta > -1/2$ and $\beta > 0$ are hyperparameters, and the densities for $\gamma$ are given by $f_\textsc{py}(\gamma) = 4^{-\theta}\Gamma(\theta + 1/2)^{-1} \gamma^{2\theta}\exp\{-\gamma^2/4\}$ and $f_\textsc{ig}(\gamma) = (\sqrt{\pi})^{-1}\exp\{\beta - \beta^2 / \gamma^2 - \gamma^2/4\}$, for the Pitman-Yor and the inverse Gaussian case, respectively. These results can be deduced from \citet{Favaro2009} and \citet{Lijoi2005, Lijoi2007b}. When $\sigma = 1/2$ the \textsc{ngg} process is also known as the normalized inverse Gaussian process \citep{Lijoi2005}.

The prior distributions in~\eqref{eq:prior_ap} are  somewhat restrictive since, by fixing the asymptotic regime to $\sigma = 1/2$, there is a single parameter ($\theta$ or $\beta$) controlling the prior mean and variance for the $\sigma$-diversity. Here, we consider a generic prior law $p(\mathrm{d}\gamma)$. The posterior for~$\gamma$ takes the form $p(\mathrm{d}\gamma \mid X_1,\dots,X_n) \propto p(\mathrm{d}\gamma) V_{n, k}(\gamma)$, where $V_{n, k}(\gamma)$ is defined in \eqref{eq:ap}. Na\"ive sampling algorithms for $p(\mathrm{d}\gamma \mid X_1,\dots,X_n)$ require evaluating the Hermite function $h_\nu(t)$, leading to numerical instabilities. For negative integers $\nu \in \{-1, -2, \dots\}$, the function $h_\nu(t)$ may be computed recursively using $h_{\nu + 1}(t) = t h_\nu(t) - \nu h_{\nu - 1}(t)$, with $h_0(t) = 1$ and $h_{-1}(t) = \Phi(t) / \phi(t)$, where $\Phi(t)$ and $\phi(t)$ are the cumulative distribution function and density of a standard Gaussian, respectively. Unfortunately, this recursion is only useful for relatively small values of $n$ and $k$. As a more robust alternative, we propose a \emph{data augmentation} strategy arising from an integral representation of Hermite polynomials: $h_\nu(t) = \Gamma(-\nu)^{-1}\int_0^\infty e^{-u^2/2 - tu}u^{-\nu - 1}\mathrm{d}u$ for any $\nu < 0$ \citep[][\S 10.5]{Lebedev1965}. We introduce a positive latent variable $U_{n,k}$ conditionally on which inference on $\gamma$ becomes straightforward, and standard sampling strategies can be employed. For $n \ge 2$ we consider the following joint likelihood for $X_1,\dots,X_n$ and $U_{n,k}$: 
\begin{equation}
\label{eq:augmentation}
\Pi(n_1,\dots,n_k, u \mid \gamma) = \frac{2^{n - k/2 - 1/2}}{\Gamma(2n - k -1)} \gamma^{k-1} u^{2n - k -2} e^{-u^2/2 - \sqrt{2}\gamma u} \prod_{j=1}^k(1/2)_{n_j-1}.
\end{equation}
It is easy to check that $\int_0^\infty \Pi(n_1,\dots,n_k, u \mid \gamma)  \mathrm{d}u = \Pi(n_1,\dots,n_k\mid \gamma)$, with the latter being the \textsc{eppf} of the Aldous-Pitman model. Although equation~\eqref{eq:augmentation} is helpful for posterior inference under any prior choice, a particularly simple Gibbs sampling algorithm is available if we let $\gamma \sim \text{Gamma}(a_\gamma, b_\gamma)$ because the corresponding full conditional densities for $\gamma$ and $U_{n,k}$ are
\begin{equation*}
f_\gamma(\gamma \mid -) \propto \gamma^{a_\gamma + k - 2}e^{- (b_\gamma + \sqrt{2}u )\gamma}, \qquad f_{U_{n,k}}(u \mid -) \propto u^{2n - k -2} e^{-(u/\sqrt{2}  + \gamma) ^2},
\end{equation*}
which means that $(\gamma \mid -) \sim \text{Gamma}(a_\gamma + k - 1, b_\gamma + \sqrt{2} u)$ is conditionally conjugate. Additionally, the conditional distribution of $(U_{n,k} \mid -)$ belongs to the family of modified half-normal distributions \citep{Sun2023}, making it straightforward to simulate due to its log-concave density \citep{Devroye1986}. Alternatively, we can circumvent the need for \textsc{mcmc}. Through simple calculus, one can derive an explicit expression for the density of $U_{n,k} \mid X_1,\dots,X_n$, which is $f_{U_{n,k}}(u \mid 
X_1,\dots,X_n) \propto u^{2n - k -1}e^{-u^2/2}(b_\gamma + \sqrt{2}u)^{-(a_\gamma + k -1)}$. This random variable is also easy to simulate, allowing for iid sampling from the posterior distribution of $\gamma \mid X_1,\dots,X_n$ in two steps: first, we sample from $U_{n,k} \mid X_1,\dots,X_n$, and then from $\gamma \mid U_{n,k}, X_1,\dots,X_n$.

The data-augmentation strategy we just proposed has further applications. In fact, it can be verified that the predictive scheme of the Aldous-Pitman process, which enables the Monte Carlo approximation of the taxon accumulation curve, can also be expressed in terms of the latent variable $U_{n, k}$. The predictive distribution is given by:
\begin{equation*}
\mathds{P}(X_{n+1} \in \cdot \mid \gamma, X_1,\dots,X_n) = \frac{\sqrt{2} \gamma}{2n - k - 1} \mathds{E}(U_{n,k}) P(\cdot) + \frac{\mathds{E}(U_{n,k}^2)}{(n - k/2)(2n - k - 1)}\sum_{j=1}^k(n_j - 1/2)\delta_{X^*_j}(\cdot),
\end{equation*}
where the expectations are taken with respect to $f_{U_{n,k}}(u \mid -) \propto u^{2n - k -2} e^{-(u/\sqrt{2}  + \gamma) ^2}$. With this representation in hand, applying some probability calculus leads to a sampling procedure for $(X_{n+1} \mid X_1,\dots,X_n)$ described in Algorithm~\ref{algo1}; see the Appendix for further details.

\section{Taxonomic Gibbs-type Priors}\label{sec:taxonomic}

\subsection{A modeling framework for taxonomic data}

In this Section, we move away from classical species sampling models described in Section~\ref{sec:gibbstype}. Here, we consider a vector of labels $\bm{X}_n = (X_{n,1},\dots,X_{n,L})$ representing multiple taxonomic levels, where each $X_{n, \ell} \in \mathds{X}_\ell$ denotes the value of the $n$th statistical unit at the $\ell$th layer of the taxonomy. This richer data structure is used to provide a multifaceted description of biodiversity.  We extend the enriched constructions of \citet{Wade2011, Rigon2025}, which are specific to $L = 2$ and limited to Dirichlet or Pitman--Yor priors. Instead, we will rely on general Gibbs-type priors. None of these works aim to infer biodiversity, which is the main goal of this paper. 

As before, we assume that $(\bm{X}_n)_{n\ge 1}$ is an infinite sequence of exchangeable observations, implying
\begin{equation}\label{eq:model_taxa}
\begin{aligned}
(X_{n,1},\dots,X_{n,L}) \mid \tilde{q} &\overset{\textup{iid}}{\sim} \tilde{q}, \qquad n \ge 1,\\
\tilde{q} &\sim \mathcal{Q}_L,
\end{aligned}
\end{equation}
where $\tilde{q}$ is a random probability measure on the product space $\mathds{X}_1 \times \cdots \times \mathds{X}_L$ and $\mathcal{Q}_L$ is its prior law. Directly using a \textsc{dp} for $\tilde{q}$ would be inappropriate as it disregards the hierarchical structure of the taxonomy. Here, we decompose $\tilde{q}$ in a Markovian fashion, that is we assume
\begin{equation}\label{eq:markovian}
\tilde{q}(\mathrm{d} x_1, \dots, \mathrm{d} x_L) = \tilde{p}_1(\mathrm{d} x_1) \tilde{p}_2(\mathrm{d} x_2 \mid x_1) \cdots \tilde{p}_L(\mathrm{d} x_L \mid x_{L-1}).
\end{equation}
Each $\tilde{p}_\ell(\cdot \mid x_{\ell-1})$ represents a conditional random probability measure on $\mathds{X}_\ell$ that depends solely on the values of the parent level $x_{\ell -1}$. %More complex dependence structure, such as considering $\tilde{p}_\ell(\cdot \mid x_1,\dots,x_{\ell - 1})$, which accounts for all values of the previous layers, would pose significant modeling challenges. 
Thus, the prior distribution for $\tilde{q}(\mathrm{d} x_1, \dots, \mathrm{d} x_L)$ is induced by choosing suitable priors for $\tilde{p}_1$ and  $\tilde{p}_\ell(\cdot \mid x_{\ell-1})$. Let $\tilde{p} \sim \text{Gibbs}(S_\sigma, \sigma; P)$ denote a Gibbs-type prior with deterministic $\sigma$-diversity $S_\sigma$ and weights $V_{n,k}(S_\sigma)$, so that $\sigma = 0$ corresponds to a Dirichlet process ($S_\sigma = \alpha$), whereas the cases $\sigma < 0 $ and $\sigma \in (0,1)$ correspond to a Dirichlet-multinomial ($S_\sigma = H$) and a $\sigma$-stable Poisson-Kingman ($S_\sigma = \gamma$), respectively. Moreover, let $P$ be a diffuse probability measure encoding the taxa. A \emph{taxonomic Gibbs-type prior} is then defined as follows
\begin{equation}\label{eq:taxonomic_Gibbs}
\tilde{p}_1 \mid S_{\sigma_1} \sim \text{Gibbs}(S_{\sigma_1}, \sigma_1; P_1), \quad \tilde{p}_\ell(\cdot \mid x_{\ell-1}) \mid S_{\sigma_\ell}(x_{\ell-1}) \overset{\text{ind}}{\sim} \text{Gibbs}(S_{\sigma_\ell}(x_{\ell-1}), \sigma_\ell; P_\ell), \quad \ell = 2,\dots,L.
\end{equation}
There are potentially infinitely many $\tilde{p}_\ell(\cdot \mid x_{\ell-1})$, that is, one for each label $x_{\ell-1}$, and they are independent among themselves. In addition, there are no shared values between observations belonging to different parents, i.e. the same \emph{species} cannot belong to multiple \emph{genera}. A crucial assumption of~\eqref{eq:taxonomic_Gibbs} is that the values $\sigma_\ell$, governing the asymptotic behavior, do not depend on $x_{\ell - 1}$, albeit they are allowed to change across layers, for $\ell = 1,\dots,L$. Thus, conditional $\sigma$-diversities $S_{\sigma_\ell}(x_{\ell - 1})$ are comparable within the same layer $\ell$, because they are expressed on the same scale. If this were not the case, the interpretation of the diversities would be problematic. 

\begin{figure}
\centering
\includegraphics[width=0.8\textwidth]{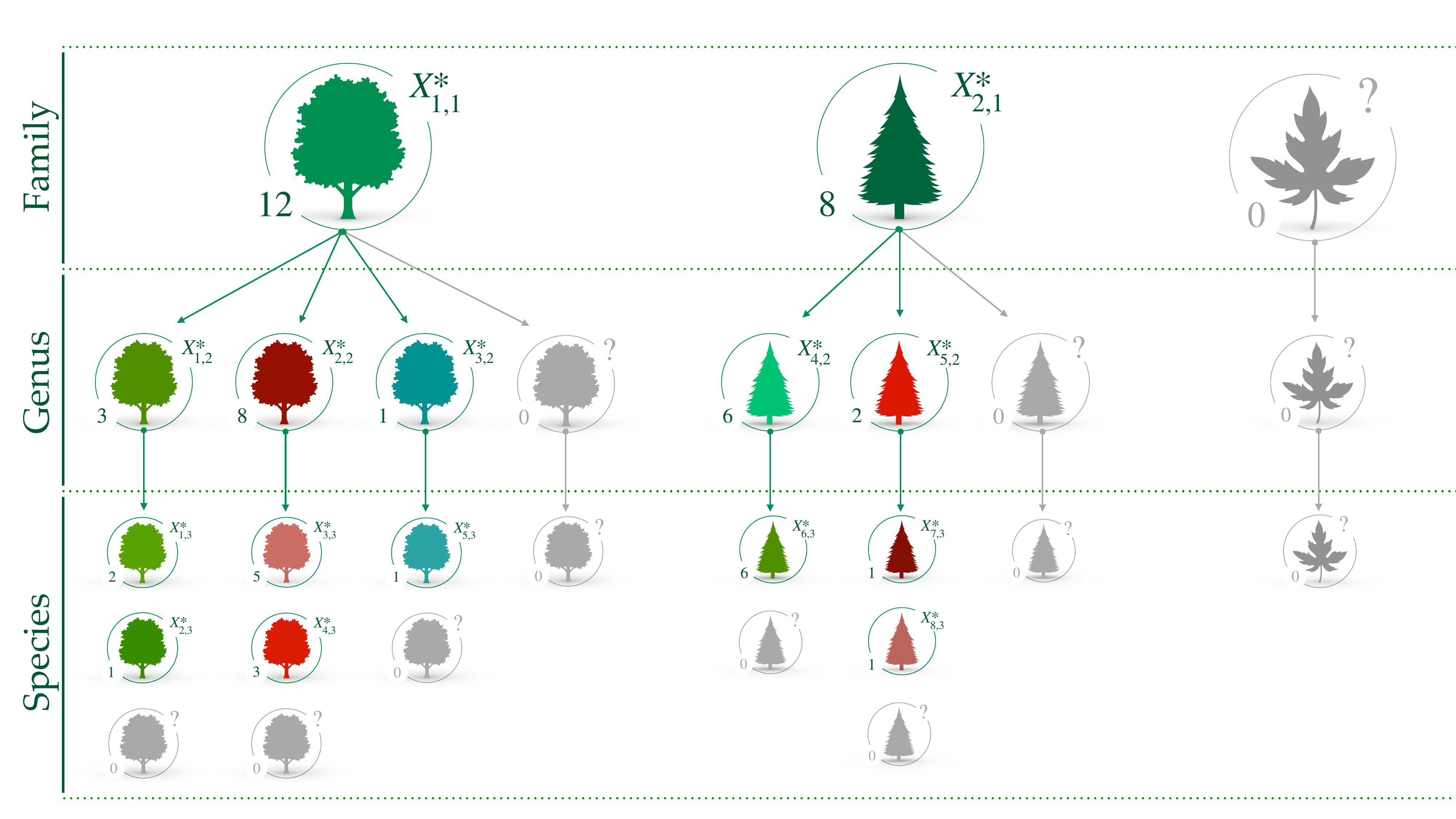}
\caption{A realization from a taxonomic Gibbs-type priors. In this example, the sample size is $n = 20$, and there are $k_1 = 2$ distinct families, $k_2 = 5$ genera and $k_3 = 8$ species. The frequencies $n_{j,\ell}$ associated to the distinct taxa $X^*_{j,\ell}$ are indicated at the bottom of each circle.  Grey circles denote the possibility of discovering new taxa. \label{fig:taxonomies}}
\end{figure}

The model outlined in equations~\eqref{eq:model_taxa}-\eqref{eq:taxonomic_Gibbs} admits a nested representation, depicted in Figure~\ref{fig:taxonomies}, which better clarifies the role of each $p_\ell(\cdot \mid x_{\ell-1})$. The observations can be sampled as follows:
\begin{equation*}
X_{n, 1} \mid \tilde{p}_1 \overset{\text{iid}}{\sim} \tilde{p}_1, \qquad X_{n, \ell} \mid X_{n, \ell-1} = x_{\ell-1}, \: \tilde{p}_\ell(\cdot \mid x_{\ell-1}) \overset{\text{ind}}{\sim} \tilde{p}_\ell(\cdot \mid x_{\ell-1}), 
\end{equation*}
for $\ell = 2,\dots,L$, and $n \ge 1$. The data of the first layer $(X_{n,1})_{n \ge 1}$ follows the same Gibbs-type mechanism that has been extensively described in Section~\ref{sec:gibbstype}. In the subsequent layers, say the $\ell$th, the observations $X_{n,\ell}$ are grouped according to parent's value $X_{n, \ell-1} = x_{\ell-1}$. The data $X_{n,\ell}$ with the same parent are iid samples from $\tilde{p}_\ell(\cdot \mid x_{\ell-1})$, which, in turn, is distributed as a Gibbs-type process with weights $V_{n,k}\{S_{\sigma_\ell}(x_{\ell-1})\}$. This representation leads to a predictive mechanism that extends~\eqref{eq:predictive}. The discreteness of each $\tilde{p}_\ell$ implies that there will be ties in the realization of $\bm{X}_1,\dots,\bm{X}_n$ with positive probability. At the $\ell$th layer, there will be $K_{n,\ell} = k_\ell$ distinct taxa with labels $X^*_{1,\ell}, \dots,X^*_{k_\ell,\ell}$ and frequencies $n_{1,\ell},\dots,n_{k_\ell,\ell}$, which may be grouped according to their parent. Let $n(x_{\ell-1})$ be the number of observations generated by the parent $x_{\ell-1}$, let $k_\ell(x_{\ell-1})$ be the number of distinct labels generated by $x_{\ell-1}$, and define the set $C(x_{\ell - 1}) = \{j : \text{ label } X^*_{j,\ell} \text{ is a child of } x_{\ell - 1} \}$, so that $k_\ell(x_{\ell-1}) = |C(x_{\ell - 1})|$ and $n(x_{\ell-1}) = \sum_{j \in C(x_{\ell-1})} n_{j, \ell}$. The predictive law is presented in the following proposition.

\begin{proposition}\label{prop:taxon_urn} Suppose $(\bm{X}_n)_{n \ge 1}$ is an exchangeable sequence as in equation~\eqref{eq:model}, with $\tilde{q}$ distributed as in~\eqref{eq:taxonomic_Gibbs}, then for any $n \ge 1$
\begin{equation*}
\mathds{P}(X_{n+1, 1} \in \cdot \mid S_{\sigma_1}, \bm{X}_1,\dots,\bm{X}_n) = \frac{V_{n+1, k_1+1}(S_{\sigma_1})}{V_{n,k_1}(S_{\sigma_1})}P_1(\cdot) + \frac{V_{n+1, k_1}(S_{\sigma_1})}{V_{n,k_1}(S_{\sigma_1})}\sum_{j=1}^{k_1}(n_{j,1} - \sigma_1)\delta_{X^*_{j,1}}(\cdot).
\end{equation*}
Moreover for $\ell = 2,\dots, L$ and $n \ge 1$
\begin{equation*}
\begin{aligned}
&\mathds{P}(X_{n+1, \ell} \in \cdot \mid X_{n+1,\ell-1} = x_{\ell-1}, S_{\sigma_\ell}(x_{\ell-1}), \bm{X}_1,\dots,\bm{X}_n)= \\
&\:=\frac{V_{n(x_{\ell-1}) + 1,k_\ell(x_{\ell-1}) + 1}\{S_{\sigma_\ell}(x_{\ell-1})\})}{V_{n(x_{\ell-1}),k_\ell(x_{\ell-1})}\{S_{\sigma_\ell}(x_{\ell-1})\}} P_\ell(\cdot) + \frac{V_{n(x_{\ell-1}) + 1,k_\ell(x_{\ell-1})}\{S_{\sigma_\ell}(x_{\ell-1})\})}{V_{n(x_{\ell-1}),k_\ell(x_{\ell-1})}\{S_{\sigma_\ell}(x_{\ell-1})\}}\sum_{j\in C(x_{\ell-1})}(n_{j,\ell} - \sigma_\ell)\delta_{X^*_{j,\ell}}(\cdot).
\end{aligned}
\end{equation*}
\end{proposition}

Proposition~\ref{prop:taxon_urn} expresses the learning mechanism in full generality, and it highlights the nested structure induced by taxonomic Gibbs-type priors. For specific choices of $\sigma_\ell$ the predictive law can be specialized by replacing the $V_{n,k}\{S_{\sigma_\ell}(x_{\ell-1})\}$'s with their specific values. For instance, if $\sigma_\ell = 0$ and $\ell = 2,\dots,L$ we get the following formula
\begin{equation*}
\begin{aligned}
&\mathds{P}(X_{n+1, \ell} \in \cdot \mid X_{n+1,\ell-1} = x_{\ell-1}, \alpha(x_{\ell-1}), \bm{X}_1,\dots,\bm{X}_n)= \\
&\qquad\qquad\qquad\qquad=\frac{\alpha(x_{\ell-1})}{n(x_{\ell-1}) + \alpha(x_{\ell-1})} P_\ell(\cdot) + \frac{1}{n(x_{\ell-1}) + \alpha(x_{\ell-1})} \sum_{j\in C(x_{\ell-1})}n_{j,\ell}\delta_{X^*_{j,\ell}}(\cdot),
\end{aligned}
\end{equation*}
which is a taxon-specific urn scheme. The diversity $\alpha(x_{\ell -1})$ refers to observations whose parent is $x_{\ell-1}$. The nested predictive mechanism described in Proposition~\ref{prop:taxon_urn} essentially states that (i) the data $X_{1,1},\dots,X_{n,1}$ in the first layer of the taxonomy follow a Gibbs-type process; (ii) the data  $X_{1,\ell},\dots,X_{n,\ell}$ for the subsequent branches follow independent predictive mechanisms that are specific to the $x_{\ell-1}$ value of the parent level. As depicted in Figure~\ref{fig:taxonomies}, if a new taxon is observed at a certain layer of the taxonomy, it leads to the discovery of new values for all the children levels. 

\subsection{Estimation of conditional biodiversity}

Compared to classical species sampling models, taxonomic Gibbs-type priors allow us to describe \emph{conditional biodiversities} $S_{\sigma_\ell}(x)$ of specific branches of the taxonomy. The first layer is modeled as a classical Gibbs-type prior, with a single $\sigma$-diversity $S_{\sigma_1}$, a case that has been extensively discussed in Section~\ref{sec:gibbstype}. Conversely, at the generic $\ell$th layer there are, potentially, infinitely many $\sigma$-diversities $S_{\sigma_\ell}(x)$, one for each value of $x \in \mathds{X}_\ell$. So far, we have treated the diversities $S_{\sigma_\ell}(x)$ as deterministic, but in practice, they must be learned from the data, and this requires careful prior elicitation. A sample $\bm{X}_1,\dots,\bm{X}_n$ provides information only about the diversities $S_{\sigma_{\ell}}(X^*_{1,\ell-1}),\dots,S_{\sigma_{\ell}}(X^*_{k_\ell,\ell-1})$ associated to the $X^*_{j, \ell-1}$ observed taxa, for $\ell = 2,\dots,L$. Moreover, from the independence of random conditional distributions $\tilde{p}_\ell(\cdot\mid x_{\ell-1})$, the likelihood function of a taxonomic Gibbs-type model factorizes as follows
\begin{equation}\label{eq:likelihood_taxonomic}
\mathscr{L}(\bm{X}_1,\dots,\bm{X}_n \mid \bm{S}) \propto V_{n,k_1}(S_{\sigma_1}) \prod_{\ell = 2}^{L} \prod_{j=1}^{k_{\ell-1}} V_{n(X^*_{j,\ell-1}),k(X^*_{j,\ell-1})}\{S_{\sigma_\ell}(X^*_{j, \ell-1})\},
\end{equation}
where $\bm{S} = (S_{\sigma_1}, S_{\sigma_2}(X^*_{1,1}), \dots, S_{\sigma_{\ell}}(X^*_{j,\ell-1}), \dots, S_{\sigma_L}(X^*_{k_{L-1},L-1}))$. %This directly follows from the definition, along the lines of \citet{Wade2014} for the enriched Dirichlet process. 
The likelihood function underlines that the data alone are not informative about the unobserved branches of the taxonomy, i.e., about $S_{\sigma_\ell}(x)$ whenever the taxon $x \in \mathds{X}_\ell$ has not been observed in the sample. 

We offer some suggestions on modeling the infinite collection of conditional diversities $S_{\sigma_\ell}(x)$. 
A simple approach assumes that the $S_{\sigma_\ell}(x)$'s are iid samples from a layer-specific prior law $\tilde{p}_\ell$, for any $x \in \mathds{X}_\ell$, independently over $\ell = 2,\dots,L$. This is computationally convenient because the posterior distributions of the conditional $\sigma$-diversities are independent due to the factorized likelihood in~\eqref{eq:likelihood_taxonomic}, and they can be inferred by applying the tools of Section~\ref{sec:gibbstype} for each $S_{\sigma_\ell}(x)$. 
However, the posterior law of $S_{\sigma_\ell}(x)$ for any unobserved taxon $x \in \mathds{X}_\ell$ coincides with its prior distribution because there is no information in the likelihood. 
Hence, it is appealing to borrow information across conditional diversities belonging to the same layer through hierarchical modelling.
%In practice, a similar problem would occur even for $ S_{\sigma_{\ell}}(X^*_{j,\ell-1})$ whenever the sample size $n(X^*_{j,\ell-1})$ of that specific branch is low. 
For example, when $\sigma_\ell = 1/2$ we may specify $S_{\sigma_\ell}(x) \mid a_\ell, b_\ell \overset{\textup{iid}}{\sim} \text{Gamma}(a_\ell, b_\ell)$, where each $(a_\ell, b_\ell)$ follows a hyperprior. Such a specification borrows strength among the diversities within the same layer $\ell$, but not across.
The factorized structure of~\eqref{eq:likelihood_taxonomic} ensures that \textsc{mcmc} algorithms can be carried out separately for each layer. We will consider an example in Section~\ref{sec:app2}. More sophisticated approaches may induce dependence across layers, for example, incorporating phylogenetic information, but we do not pursue this here. 

\begin{remark} Marginal and conditional $\sigma$-diversities provide two different, but complementary, summaries of a complex phenomenon. Conditional diversities strongly depend on the chosen taxonomical structure. Potentially the models developed in this article can be used in concert with models characterizing variation in genetic sequences and morphology data for automatic taxonomic classification of samples, allowing for the discovery of new taxa, in a related manner to \citet{Zito2023c}. 

%as certain layers may be included in one analysis and omitted in another. 
%These structural choices, including the number of layers $L$ to be considered, are highly relevant but do not have a statistical answer. Rather, they reflect the focus of the ecologist on certain scientific aspects. 
\end{remark}

\section{The tree flora in the Amazonian Basin}
\subsection{The estimation of $\sigma$-diversity}\label{sec:app1}

We consider the comprehensive tree data set from the Amazon Basin and the Guiana Shield (Amazonia) provided by \citet{TerSteege2013}, which is openly accessible online. This dataset integrates multiple sources and includes $1170$ sampling plots well distributed among regions and forest types of the Amazon Basin, whose locations are shown in Figure~\ref{fig:sites}. The immense size of the Amazon Basin has limited research on its tree communities at local and regional levels. Scientists still lack a complete understanding of the number of species in the Amazon, as well as their abundance and rarity \citep{Hubbell2008, TerSteege2013}, which leaves the world’s largest tropical carbon reserve mostly unknown to ecologists. The plots included a total of $k = 4,962$ tree species, comprising $747$ genera and $115$ families, based on $n = 553,949$ observations. In \citet{TerSteege2013}, Fisher’s log-series model was applied to estimate the total number of species, generating approximately $15,000$ tree species. About $200$ of these species were classified as \emph{hyperdominant}, which means they are so common that, collectively, they represent half of all trees in the Amazon. These estimates suggest that there may be over $10,000$ rare and poorly known species in the Amazon that could be at risk. We aim to provide a more refined statistical analysis of these data based on a Bayesian nonparametric analysis of the $\sigma$-diversity. In practice, our model-based approach allows us to: (i) check the validity of the underlying assumptions of the model; (ii) provide formal tools for uncertainty quantification even in the presence of model misspecification; (iii) tie together multiple aspects of the analysis within a unified model, making it clear and coherent. 

\begin{figure}[tbp]
\centering
    \includegraphics[width=0.7\textwidth]{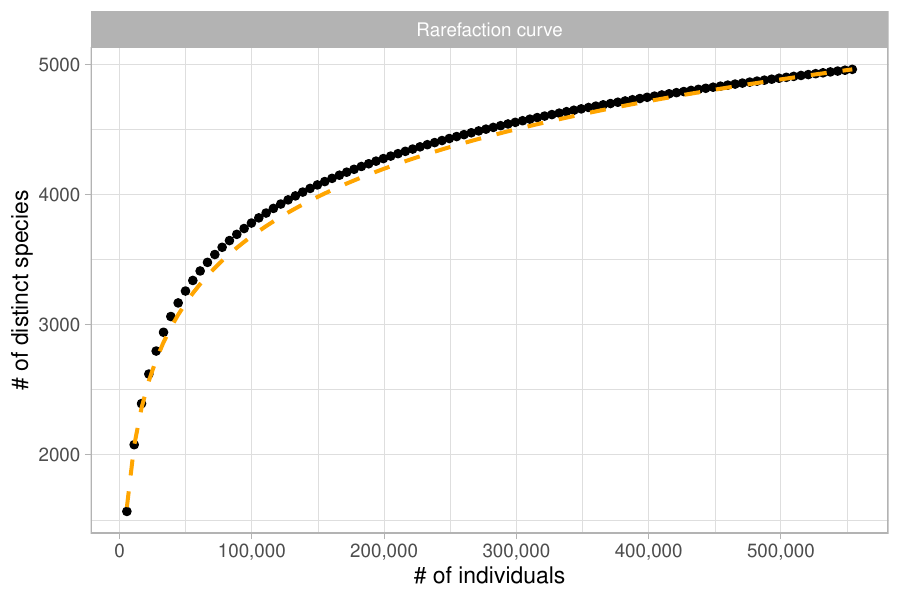}
    \caption{Rarefaction curves for the Amazonian tree species dataset from \citet{TerSteege2013}. Black dots represent the ``classical rarefaction'' values $\bar{K}_1, \dots, \bar{K}_n$, as described in Section~\ref{sec:check}; for clarity, only a subset of points is shown. The orange dashed line represents the model-based rarefaction $\mathds{E}(K_n \mid \hat{\alpha}_\textsc{ml}) = \sum_{i=1}^n \hat{\alpha}_\textsc{ml} / (\hat{\alpha}_\textsc{ml} + i - 1)$ from a Dirichlet process model.}
    \label{fig:rarefaction}
\end{figure}

\begin{figure}[tbp]
\centering
    \includegraphics[width=0.7\textwidth]{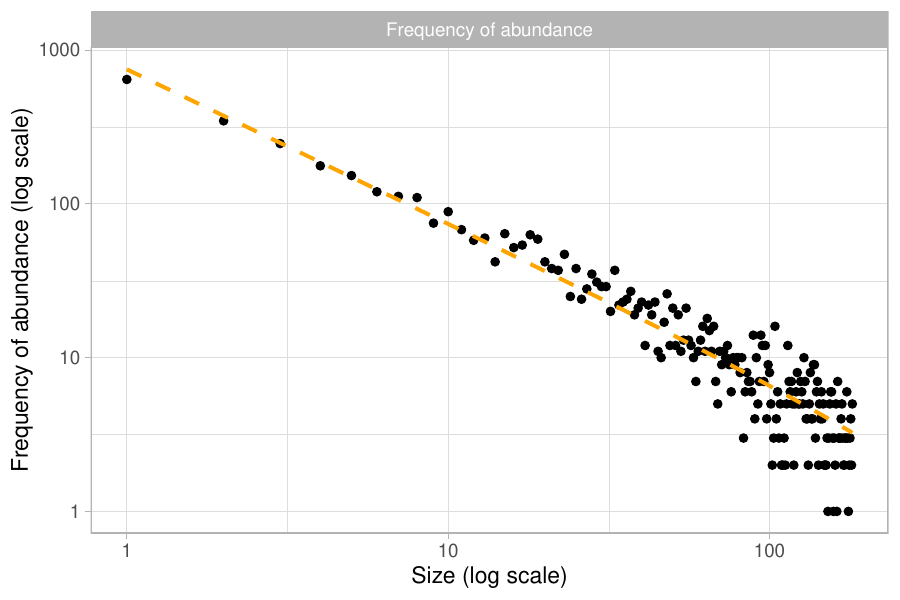}
    \caption{Frequency of abundance plot for the Amazonian tree dataset from \citet{TerSteege2013}. Black dots represent the observed frequency counts $m_r$ for sizes $r = 1, \dots, n$, where $m_r$ denotes the number of tree species that appear with frequency $r$ in the observed sample. The orange dashed line shows the model-based expectations $\mathds{E}(M_{r,n} \mid \hat{\alpha}_\textsc{ml})$ for $r = 1, \dots, n$, with the formula provided in Section~\ref{sec:check}. Both axes are displayed on a log scale for clarity.}
    \label{fig:freq_of_freq}
\end{figure}

\begin{figure}[tbp]
\centering
    \includegraphics[width=0.7\textwidth]{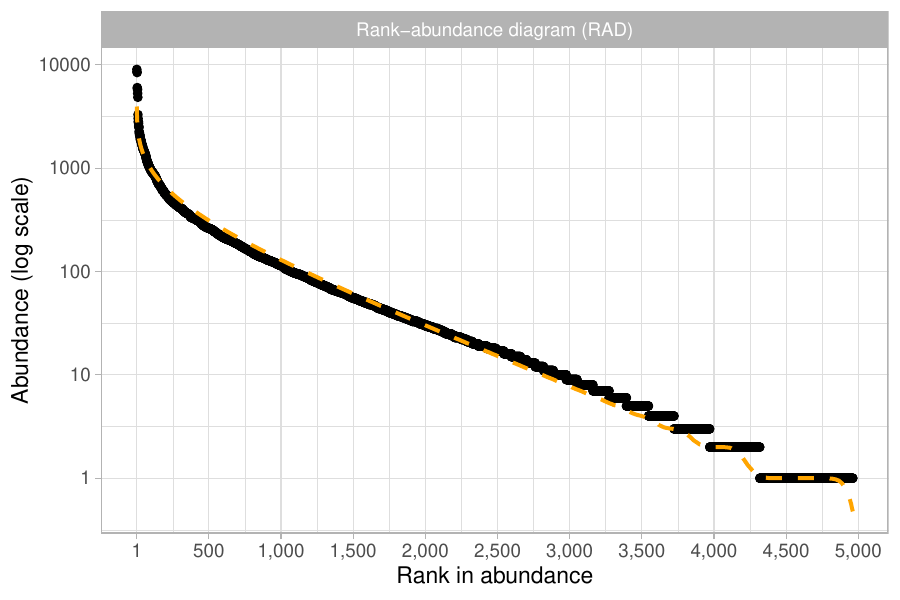}
    \caption{Rank in abundance diagram (\textsc{rad}) for the Amazonian tree dataset from \citet{TerSteege2013}. Black dots represent the observed abundances $n_1,\dots,n_k$, ranked from the most common to the least common, in log scale. The dashed orange line represents the expected abundances from a Dirichlet process model with parameter $\hat{\alpha}_\textsc{ml}$, ranked from the most common to the least common. Model-based values are obtained via Monte Carlo by sampling from the \citet{Blackwell1973}  urn scheme with parameter $\hat{\alpha}_\textsc{ml}$ and averaging the ranked abundances over $1,000$ replicates.}
    \label{fig:RAD}
\end{figure}

The observations $X_1, \dots, X_n$ represent tree species and are assumed to be \emph{exchangeable} under a Gibbs-type species sampling model. In particular, we will show that it is reasonable to assume $\sigma = 0$, leading to the following hierarchical model:
\begin{equation*}
\begin{aligned}
    X_i \mid \tilde{p} &\overset{\text{iid}}{\sim} \tilde{p}, \qquad &&n \ge 1, \\
    \tilde{p} \mid \alpha &\sim \textsc{dp}(\alpha P), \qquad &&\alpha \sim p(\mathrm{d}\alpha).
    \end{aligned}
\end{equation*}
Recall that this specification is equivalent to Fisher’s log-series model. As a preliminary step, we obtain a maximum likelihood estimate for the fundamental biodiversity number of $\hat{\alpha}_{\textsc{ml}} = 751.23$ and a Fisher’s estimate of $\hat{\alpha}_\textsc{f} = 751.32$. The former was calculated using the \texttt{BNPvegan} R package \citep{Zito2023b}, while the latter was obtained through the \texttt{vegan} R package \citep{vegan2024}. As previously mentioned, these two estimates are asymptotically equivalent and nearly indistinguishable for large $n$. To assess the validity of the Dirichlet process specification we follow the guidelines outlined in Section~\ref{sec:check} and we consider the quantities $\mathds{E}(K_n \mid \hat{\alpha}_\textsc{ml})$ and $\mathds{E}(M_{r,n} \mid \hat{\alpha}_\textsc{ml})$. We then compare these expectations with their empirical counterparts, as displayed in Figure~\ref{fig:rarefaction} (rarefaction) and Figure~\ref{fig:freq_of_freq} (frequency of abundance). Both plots give similar conclusions: although the fit is not perfect, the predictions are remarkably close to the observed data. %, which is surprising given that the Dirichlet process has a single parameter $\alpha$. 
Indeed, the growth rate for the number of distinct species in Figure~\ref{fig:rarefaction} looks roughly logarithmic, supporting the choice $\sigma = 0$. At the same time, there are mild discrepancies between the data and the model predictions: the empirical rarefaction curve has a slightly higher curvature than the model fit. Moreover, Figure~\ref{fig:freq_of_freq} shows that the Dirichlet process model slightly overestimates the number of singletons, and in particular, we get $\mathds{E}(M_{1, n} \mid \hat{\alpha}_\textsc{ml}) = 750.22$ and $m_1 = 645$. %Recall that for large $n$ we have $\mathds{E}(M_{1, n} \mid \alpha) \approx \alpha$. 
Finally, we show in Figure~\ref{fig:RAD} the rank in abundance diagram (\textsc{rad}), which is the main graphical tool considered in \citet{TerSteege2013} to validate the adequacy of the Fisher's log-series model. Once again, we conclude that the fit is quite good. 
%As a general remark, we do not expect the fit of a Dirichlet process model to be this good in other case studies nor the choice $\sigma = 0$ to be always appropriate; for instance, polynomial growths are typically observed in genetics \citep{Lijoi2007}. 

Despite its good predictive performance, the Dirichlet process is likely misspecified in this context. The data originates from multiple studies, each examining  different regions and forests, which makes the exchangeability assumption for $(X_n)_{n \ge 1}$ unlikely to hold. Supporting this concern, classical nonparametric and frequentist estimators such as those by \citet{Chao1984}, \citet{Chao1992}, and \citet{Chao2002} yield predictions--and associated confidence intervals--of between 5,000 and 6,000 for the total number of distinct species. These values were computed using the \texttt{SPECIES} R package \citep{Wang2011}.  However, \citet{TerSteege2013} criticized these predictions, labeling them as ``severe underestimations'', due to sharp disagreement with previous studies and expert assessments. This failure of standard nonparametric methods is not unusual in ecological studies \citep{Brose2003}, where assumptions often do not hold.  Such estimators typically perform well in localized settings, but here we are considering the entire Amazon Basin. In contrast, the inherent ``rigidity'' of the Dirichlet process model enforces a logarithmic growth rate, which helps to avoid overfitting to the observed data, thus potentially mitigating the effects of misspecification.

With these considerations in mind, we proceed to conduct a full Bayesian analysis, accounting for potential model misspecification through the use of coarsened posteriors \citep{Miller2019}. In practice, this is approximately achieved by raising the likelihood function to a factor $\rho \in (0, 1]$, which effectively deflates the sample size and increases the uncertainty; taking $\rho = 1$ recovers the standard posterior. For our analysis, the coarsened posterior distribution for $\alpha$ becomes:
\begin{equation*}
    p_\rho(\mathrm{d}\alpha \mid X_1,\dots,X_n) \propto p(\mathrm{d}\alpha) \left[\frac{\alpha^k}{(\alpha)_n}\right]^\rho.
\end{equation*}
A natural prior choice is $\alpha \sim \textsc{sg}(a, b, n)$,  the Stirling-gamma distribution of \citet{Zito2024} with parameters $(a, b, n)$, where $1 \le a/b \le n$ and  $a,b > 0$. This prior interacts well with coarsened posteriors, remaining conjugate even after tempering the likelihood. Specifically, under a Stirling-gamma prior, the coarsened posterior is given by $p_\rho(\mathrm{d}\alpha \mid X_1,\dots,X_n) \propto \alpha^{a + \rho k} / \{(\alpha)_n\}^{b + \rho}$, resulting in:
\begin{equation*}
(\alpha \mid X_1,\dots, X_n) \sim \textsc{sg}(a + \rho k, b + \rho, n).
\end{equation*}
Here, $a/b$ serves as the location of the prior distribution,while $b$ acts as a precision \citep{Zito2024}. In particular, we have $\mathds{E}(K_n) = a/b$, the prior estimate of the number of distinct species in a sample of size $n$. In our analysis, we set $a/b = 5,000$ and $b = 0.002$, implying $a = 1$, thus providing a non-informative prior centered on a plausible guess. Thanks to conjugacy, i.i.d. sampling from the coarsened posterior is straightforward through the algorithm of \citet{Zito2024}.

\begin{table}[ht]
\centering
\begin{tabular}{lr|rrrrrrr}
  \toprule
  $\rho$ & Equivalent sample size ($n \rho $) & 1\% & 25\% & 50\% & Mean & 75\% & 99\% \\ 
   \midrule
  \multicolumn{8}{l}{\emph{Fundamental biodiversity number $\alpha$}} \\
 1 & 553,949 & 725 & 743 & 751 & 751 & 759 & 779 \\ 
 0.25 & 138,487 & 699 & 736 & 751 & 751 & 767 & 806 \\ 
 0.1 & 55,395 & 669 & 726 & 751 & 751 & 776 & 839 \\ 
 0.01 & 5,539 & 514 & 673 & 747 & 753 & 827 & 1,048 \\ 
 0.001 & 554 & 208 & 517 & 713 & 766 & 956 & 1,792 \\ 
 \midrule
  \multicolumn{8}{l}{\emph{Total number of tree species $K_N$}} \\
  %\midrule
1 & 553,949 & 14,378 & 14,841 & 15,065 & 15,051 & 15,267 & 15,678 \\ 
0.25 & 138,487 & 14,139 & 14,777 & 15,052 & 15,052 & 15,327 & 15,976 \\ 
0.1 & 55,395 & 13,814 & 14,675 & 15,045 & 15,054 & 15,422 & 16,371 \\ 
0.01 & 5,539 & 11,824 & 13,981 & 14,990 & 15,077 & 16,077 & 19,097 \\ 
0.001 & 554 & 7752 & 11,906 & 14,533 & 15,246 & 17,800 & 29,058 \\ 
   \bottomrule
\end{tabular}
\caption{Posterior mean and quantiles for $\alpha$ and the total number of tree species $K_N$, for various choices of the coarsening parameter $\rho$, for the Amazonian tree dataset from \citet{TerSteege2013}. These values are based on $10^6$ Monte Carlo replicates using the Stirling-gamma sampling algorithm of \citet{Zito2024} for $\alpha$ and a Poisson approximation for $K_N$ (see the main text). \label{tab:total_number_species}}
\end{table}

We report the results in Table~\ref{tab:total_number_species}, which shows posterior means and quantiles for $\alpha$ under various choices of the tempering parameter $\rho \in \{0.001, 0.01, 0.25, 1\}$. Choosing $\rho$ is challenging, as it requires accurately quantifying the degree of misspecification. Table~\ref{tab:total_number_species} also includes the equivalent sample size $n \rho$; we note that, except for the extreme case of $\rho = 0.001$, the posterior means and medians remain stable, approximately matching the maximum likelihood estimate of $751$. The value of $\rho$ has a substantial impact on posterior variability, with smaller $\rho$ values representing conservative choices by leading to greater uncertainty. Assuming the model is correctly specified, the $98\%$ credible interval for $\alpha$ is $(725, 779)$. However, with a moderate level of coarsening, say $\rho = 0.01$, the 98\% credible interval broadens to $(514, 1048)$. Notably, the independent estimate $\hat{\alpha} = 743$ for the Amazon Basin from \citet{Hubbell2008} lies within both intervals. We provide graphical support for $\rho = 0.01$ in the Supplementary Material, using an informal elbow rule to identify the smallest $\rho$ that does not lead to strong incompatibilities with the observed data, as measured by the likelihood function \citep{Miller2019}. Figure~\ref{fig:post_alpha} displays the coarsened posterior for $\alpha$ with $\rho = 0.01$. The posterior distribution of $\alpha$, the fundamental biodiversity number, encapsulates various aspects of the data. For example, for large $n$, the expected number of singletons is $\mathds{E}(M_{1, n} \mid \alpha) \approx \alpha$, so the interval $(514, 1048)$ serves also as an estimate for the number of singletons. This also supports the choice $\rho = 0.01$, given that the observed number of singletons is $m_1 = 645$, which lies well outside the credible interval of the standard posterior. As detailed in Section~\ref{sec:diversities}, other biodiversity measures can be derived from the posterior distribution of $\alpha$. For instance, the posterior mean for the Simpson diversity, $\mathds{P}(X_1 = X_2) = 1/(1+\alpha)$, is $0.00136$, while the posterior mean for Shannon diversity is $7.1884$.

\begin{figure}[tbp]
\centering
    \includegraphics[width=0.7\textwidth]{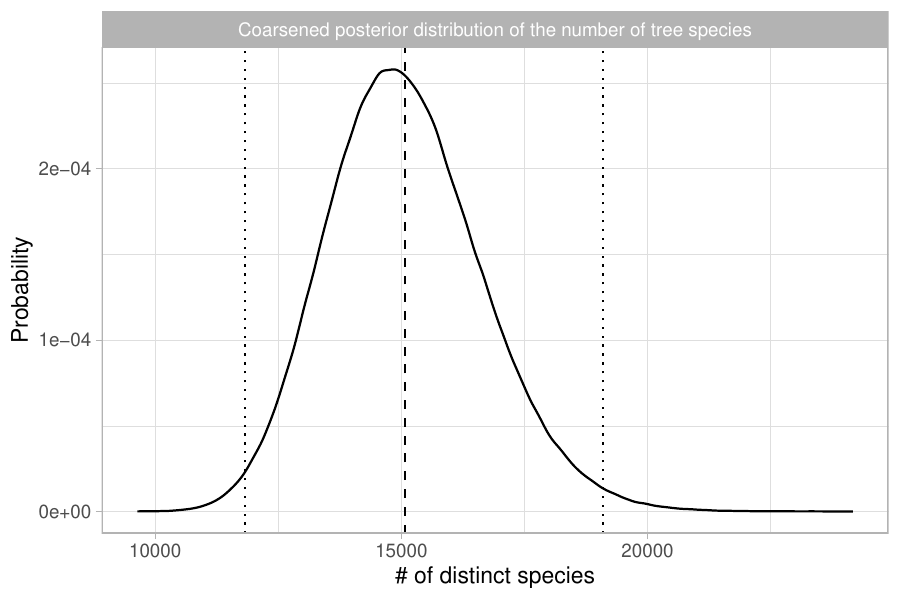}
    \caption{Coarsened posterior distribution ($\rho = 0.01$) of the total number of distinct tree species $K_N$, using the Amazonian tree dataset of \citet{TerSteege2013}. The dotted lines represent $98\%$ credible intervals. The dashed line is the posterior mean. These values are based on $10^6$ Monte Carlo replicates using the Stirling-gamma sampling algorithm of \citet{Zito2024} combined with a Poisson approximation for $K_N$ (see the main text).}
    \label{fig:total_species_coarsened}
\end{figure}

The most interesting quantity we wish to estimate is arguably the total number of tree species in the Amazon Basin. In \citet{TerSteege2013}, this is estimated using Fisher's log series model in a relatively ad-hoc manner, by extrapolating the \textsc{rad} shown in Figure~\ref{fig:RAD} to obtain approximately 15,000 to 16,000 tree species, depending on the extrapolation method used. Here, we present a more mathematically rigorous approach that also properly quantifies the associated uncertainty. Let us first recall that in the Dirichlet process model, $K_n$ diverges as $n$ grows. However, this should not be viewed as a shortcoming of the model. The population size $N$, representing the total number of trees in the Amazon Basin, is finite, implying that the total number of tree species is $K_N$. Furthermore, \citet{TerSteege2013} provides a reliable estimate of the population size, $\hat{N} = 3.949\times 10^{11}$, based on observed tree density. To be robust against potential errors in this estimate, we specify a prior for $N$ centered on $\hat{N}$ by letting $N \sim \text{Uniform}(0.5\hat{N}, 1.5\hat{N})$. This prior choice is roughly consistent with the standard errors for $\hat{N}$ given in \citet{TerSteege2013}. Note we cannot learn $N$ based on the publicly available data, so we rely on external information for this inferential step. In summary, the coarsened posterior distribution for the total number of tree species can be obtained through the following sampling steps:
\begin{equation*}
(K_N \mid \alpha, N, X_1,\dots,X_n) \sim p_K, \:\: (\alpha \mid X_1,\dots, X_n) \sim \textsc{sg}(a + \rho k, b + \rho, n), \:\: N \sim \text{Uniform}(0.5\hat{N}, 1.5\hat{N}).
\end{equation*}
In Section~\ref{sec:gibbstype}, we described general formulas for the law $p_K$, which could, in principle, be applied here. Additionally, note that the posterior expectation of $K_N$ is $\mathds{E}(K_N\mid \alpha, N, X_1,\dots,X_n) = k + \sum_{i=1}^{N - n} \alpha/(\alpha + n + i -1)$. For simplicity and computational efficiency, we rely on a well-known, highly accurate Poisson approximation valid for large values of $N$:
\begin{equation*}
(K_m^{(n)} \mid \alpha, N, X_1,\dots,X_n) \overset{.}{\sim} \text{Poisson}\left(\sum_{i=1}^{N - n} \frac{\alpha}{\alpha + n + i -1}\right),
\end{equation*}
where $\overset{.}{\sim}$ means ``approximately distributed as'', recalling that, given the data, we have $K_N = k + K_m^{(n)}$. This approximation can be rigorously justified in terms of total variation convergence, which holds for large values of $N$; see Proposition 4.8 in \citet{Ghosal2017}. We report the results in Table~\ref{tab:total_number_species}, which shows posterior means and quantiles for $K_N$ under various choices of the tempering parameter. Figure~\ref{fig:total_species_coarsened} displays the coarsened posterior of $K_N$ with $\rho = 0.01$. Our analysis suggests that the Bayesian estimate for the total number of tree species is approximately 15,000, consistent with the main finding of \citet{TerSteege2013}. Additionally, we estimate a 98\% credible interval for this total to be between 11,800 and 19,100 when $\rho = 0.01$. Notably, due to the logarithmic growth rate, the uncertainty in the posterior distribution of $K_N$ is primarily influenced by the variability in $\alpha$, rather than potential misspecifications of the population size $N$.

\subsection{Taxonomic $\sigma$-diversity}\label{sec:app2}

We now describe a more nuanced analysis of the Amazon tree dataset that incorporates the taxonomic structure of the data. This analysis uses the taxonomic Gibbs-type priors introduced in Section~\ref{sec:taxonomic} with $L = 3$, where the exchangeable observations $\bm{X}_1,\dots,\bm{X}_n$ are triplets, denoted $\bm{X}_i = (X_{i,1}, X_{i,2}, X_{i,3})$, whose elements represent the family, genus, and species of the $i$th tree, respectively. The last element of each triplet, $X_{i,3}$, corresponds to the previously analyzed tree species. We let the family labels $X_{n, 1} \mid \tilde{p}_1$ be iid samples from $\tilde{p}_1$, where $\tilde{p}_1 \sim \textsc{dp}(\alpha_1 P_1)$ for $n \ge 1$. Moreover, for the genera and species, we assume the observations as generated as follows
\begin{equation*}
X_{n, 2} \mid X_{n, 1} = x_1, \: \tilde{p}_2(\cdot \mid x_1) \overset{\text{ind}}{\sim} \tilde{p}_2(\cdot \mid x_1), \quad X_{n, 3} \mid X_{n, 2} = x_2, \: \tilde{p}_3(\cdot \mid x_2) \overset{\text{ind}}{\sim} \tilde{p}_3(\cdot \mid x_2), \quad n \ge 1.
\end{equation*}
We specify different priors for the second and third layers of the taxonomy. Denoting by $\alpha = \{\alpha(x)\}_{x \in \mathds{X}_1}$ and $\gamma = \{\gamma(x)\}_{x \in \mathds{X}_2}$, we assume
\begin{equation*}
\begin{aligned}
&\tilde{p}_2(\cdot \mid x_1) \mid \alpha(x_1) \overset{\text{ind}}{\sim} \textsc{dp}(\alpha(x_1)P(\cdot\mid x_1)), \qquad &&\tilde{p}_3(\cdot \mid x_2) \mid \gamma(x_2) \overset{\text{ind}}{\sim} \textsc{ap}(\gamma(x_2)P(\cdot\mid x_2)), \\
&\alpha \sim \mathcal{Q}_\alpha, \qquad &&\gamma \sim \mathcal{Q}_\gamma.
\end{aligned}
\end{equation*}
In other words, we assume a logarithmic growth rate for genera within families and a faster polynomial growth rate for species within each genus, induced by the Aldous-Pitman model. This taxonomic approach allows us to compare the genus-level biodiversity within each family, denoted by $\alpha(x)$, as well as the species-level biodiversity within each genus, denoted by $\gamma(x)$. To get Bayesian estimates for $\alpha(x)$ and $\gamma(x)$ we need to specify priors $\mathcal{Q}_\alpha$ and $\mathcal{Q}_\gamma$. In the former case, we let $\alpha(x) \overset{\textup{iid}}{\sim} \textsc{sg}(a_\alpha, b_\alpha, m)$, with $a_\alpha/b_\alpha = 3$, $b_\alpha = 0.1$, $m = 100$, a fairly informative prior according to which we expect, a priori and on average, 3 distinct genera for each family, in a hypothetical sample from the same family of size $m = 100$. On the other hand, we assume a hierarchical prior for $\gamma(x)$, to borrow strength across species-level biodiversities, which is needed due to the sparsity of the data. Specifically, we let
\begin{equation*}
\gamma(x) \mid a_\gamma, b_\gamma \overset{\textup{iid}}{\sim}\text{Gamma}(a_\gamma, b_\gamma), \qquad (\log{a_\gamma},\log{b_\gamma}) \sim \text{Normal}_2(\bm{\mu}_\gamma, \sigma^2_\gamma I_2).
\end{equation*}
where we set $\bm{\mu}_\gamma = (0, 0)$ and $\sigma^2_\gamma = 10^2$. 

We ran an \textsc{mcmc} algorithm for 10,000 iterations, discarding the first 1,000 as burn-in; the computational details are provided in the Supplementary Material. The sampling algorithm is particularly straightforward due to the factorized likelihood~\eqref{eq:likelihood_taxonomic}. This, combined with suitable priors $\mathcal{Q}\alpha$ and $\mathcal{Q}\gamma$, results in separate blocks of the model that can be estimated independently. The posterior distribution for the most relevant part of the model--the one describing species-within-genera biodiversity--has been coarsened with $\rho = 0.25$ to address potential misspecification. As before, this choice of coarsening is supported by an informal elbow rule.

\begin{figure}[tbp]
\centering
    \includegraphics[width=0.7\textwidth]{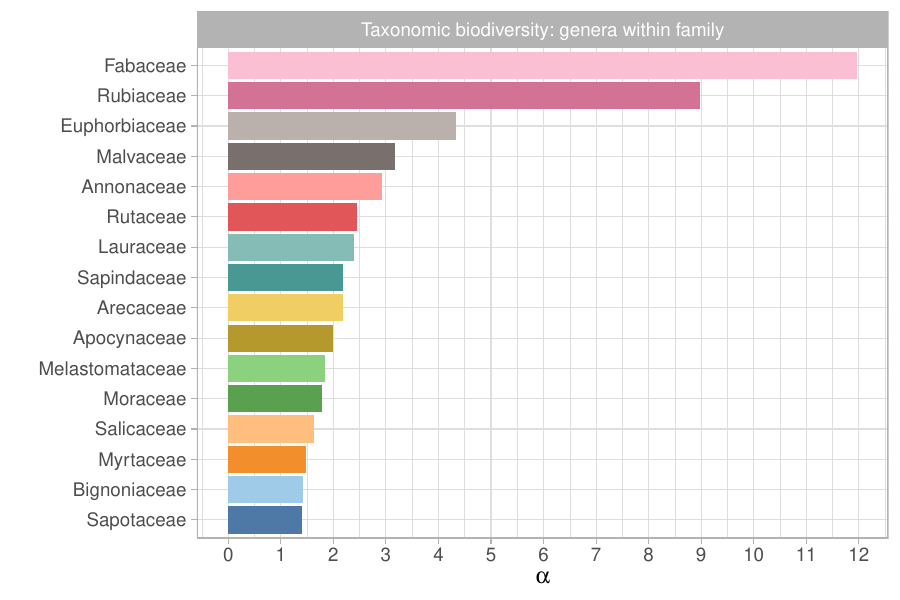}
    \caption{Taxon biodiversity: genera within family. We plot the posterior expectations $\mathds{E}(\alpha(x) \mid \bm{X}_1,\dots,\bm{X}_n)$, approximated via \textsc{mcmc}, of the $16$ most diverse families out of $115$ as measured by $\alpha(x)$. }
    \label{fig:taxonomic_family}
\end{figure}

\begin{figure}[p]
\centering
    \includegraphics[width=0.8\textwidth]{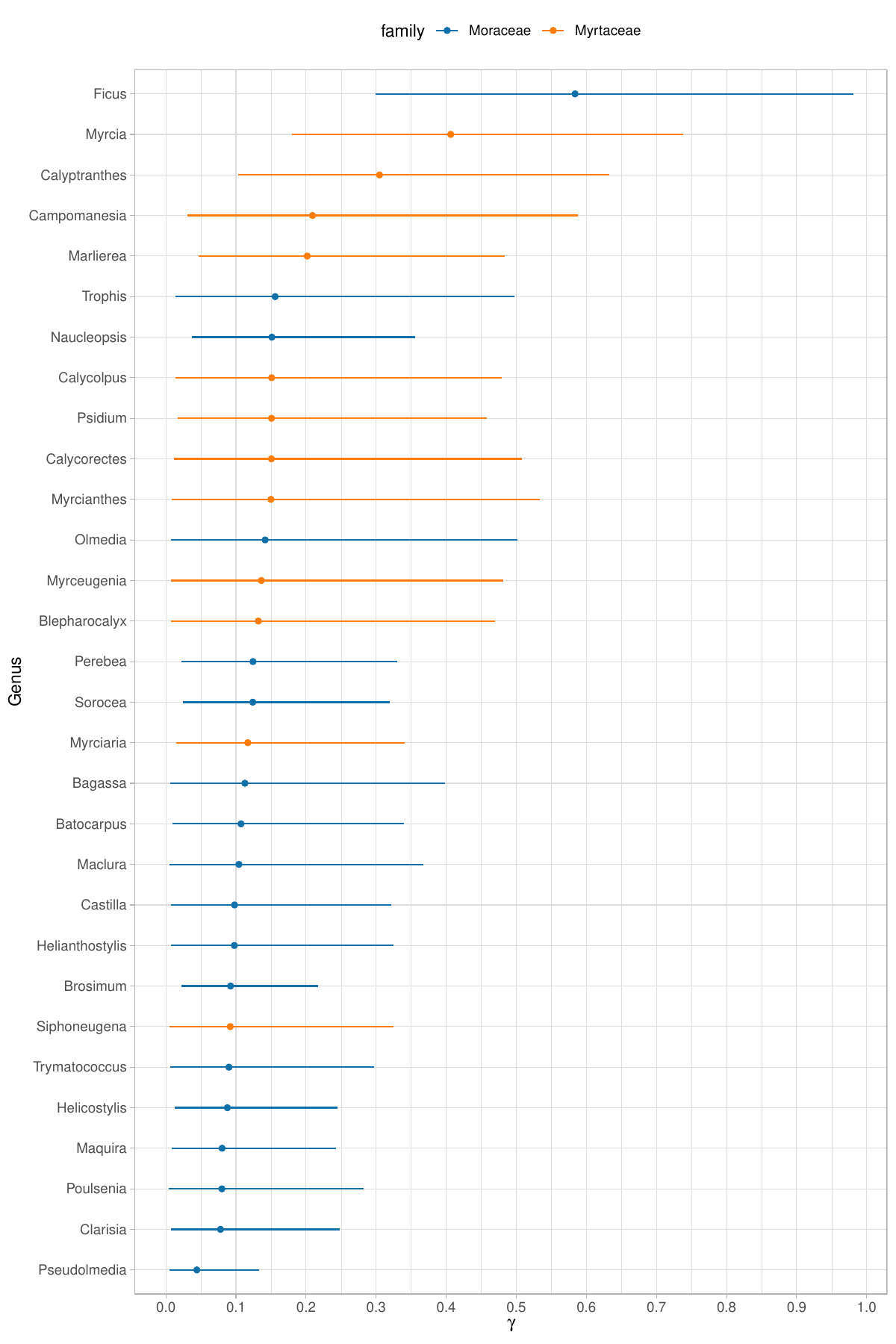}
    \caption{Taxon biodiversity: species within genus, for two randomly selected families (\emph{Moraceae}, \emph{Myrtaceae}. Dots represent the posterior means $\mathds{E}(\gamma(x) \mid \bm{X}_1,\dots,\bm{X}_n)$, while lines represent a 98\% credible interval, approximated via \textsc{mcmc}.}
    \label{fig:taxonomic_genus_diversity}
\end{figure}

We report in Figure~\ref{fig:taxonomic_family} the posterior expectations $\mathds{E}(\alpha(x) \mid \bm{X}_1,\dots,\bm{X}_n)$ for selected families. Specifically, we show the 16 most diverse of 115 families, as measured by the fundamental biodiversity numbers $\alpha(x)$. These values represent family-specific biodiversity that accounts for the variety of genera within the same taxonomic branch. However, it is important to note that $\alpha(x)$ does not reflect the number of species within each genus. The most biodiverse family is \emph{Fabaceae}, which is expected, since it is also the most abundant and rich family in the dataset. Furthermore, as observed in \citet{TerSteege2013}, \emph{Rubiaceae}--the second most diverse family according to $\alpha(x)$--has relatively few hyperdominant species, which aligns with its high diversity number. Moreover, in Figure~\ref{fig:taxonomic_genus_diversity}, we plot the posterior distributions of select genus-specific biodiversities $\gamma(x)$ for all genera in two randomly chosen families, \emph{Moraceae} and \emph{Myrtaceae}. As the plot suggests, with few exceptions, there is substantial uncertainty in each $\gamma(x)$, though this is partially reduced by borrowing strength across genera. In particular, in our hierarchical model, the estimated average biodiversity is $\mathds{E}(a_\gamma / b_\gamma \mid \bm{X}_1,\dots,\bm{X}_n) = 0.15$ with a standard deviation of $\mathds{E}(\sqrt{a_\gamma / b_\gamma^2} \mid \bm{X}_1,\dots,\bm{X}_n) = 0.11$. Two genera that stand out from the average behavior are \emph{Ficus} and \emph{Pseudolmedia}. The former ranks among the most diverse genera of trees, as evidenced by the extensive list of species documented in Plants of the World Online (\url{https://powo.science.kew.org}); see \citet{Baker2022}. To date, there are almost a thousand accepted species of \emph{Ficus} spread across Amazonia and other regions of the world. In contrast, \emph{Pseudolmedia} has only 11 known species globally (four of them observed in our dataset), all situated in the Amazon basin and adjacent areas; some of these species may be at risk of extinction.

\section{Discussion}

In this paper, we discuss what we believe to be one of the most natural definitions of biodiversity from a Bayesian perspective, unifying accumulation curves, and most existing biodiversity measures under a model-based approach. Mathematically, this approach takes advantage of Gibbs-type priors as defined in \citet{Gnedin2005}, although most of the inferential results crucial for species discovery were developed in a subsequent paper by \citet{Lijoi2007}. This theory has significant connections with ideas developed over the years, particularly by \citet{Fisher1943} and \citet{Hubbell2001}. A key natural question is understanding generalizations to different sampling mechanisms, such as those where only species presence is recorded (i.e. incidence data). This has recently been addressed in \citet{Ghilotti2024}, which introduced a broad class of models sharing structural properties with Gibbs-type priors and leads to a natural notion of biodiversity for incidence data.

We also envision several promising research directions. First, we observe that a systematic comparison of priors for $H$, representing species richness in a Dirichlet-multinomial model, is currently lacking. This is a straightforward next step that could be effectively explored with moderate effort, at least from an empirical standpoint. A second and more complex issue involves the inclusion of covariates. Estimating localized biodiversity metrics for specific regions or tracking biodiversity changes over time is both more interesting and realistic than estimating a single biodiversity number. However, directly applying covariate-dependent Dirichlet processes \citep[e.g.,][]{Quintana2022} appears unsuitable and requires careful modifications. The work of \citet{Zito2023c} goes in that direction by incorporating DNA sequencing into the model, but it does not provide synthetic measures of biodiversity. A third relevant issue refers to a clear definition of the so-called ``beta'' diversity, namely the heterogeneity of species across different sampling regions. Recently, there have been sensible developments in models that can account for shared species, most notably the hierarchical constructions described in \citet{Camerlenghi2017}, \citet{Camerlenghi2019b}. However, some computational challenges remain open, as well as a precise notion of ``beta'' diversity which should be as clear and unifying as the $\sigma$-diversity described in this paper. 

\section*{Acknowledgments}
Tommaso Rigon acknowledges support of MUR - Prin 2022 - Grant no. 2022CLTYP4, funded by the European Union - Next Generation EU. This work was partially supported by the European Research
Council under the European Union’s Horizon 2020 research and innovation programme
(grant agreement No 856506). 
\bibliographystyle{chicago}
\bibliography{biblio}

\appendix

\section{Proofs}

\subsection{Rarefaction and extrapolation curve of a Dirichlet-multinomial model}

We present here the proof of the results in Example~\ref{ex:dm}, concerning the expectations $\mathds{E}(K_n \mid H)$ and $\mathds{E}(K_m^{(n)} \mid H, X_1, \dots, X_n)$ in the Dirichlet multinomial model. To begin, let us recall that the process is defined as  $\tilde{p} = \sum_{h=1}^H W_h \delta_{Z_h}$, where $(W_1, \dots, W_H) \sim \textup{Dir}(|\sigma|, \dots, |\sigma|)$. The a priori expected value of the number of distinct values is well known \citep[e.g. ][ Chap. 3]{Pitman2006}, but we provide here a simple and alternative proof that does not require combinatorial calculus. Specifically, we have:
\begin{equation*}
\mathds{E}(K_n \mid H) = \mathds{E}\{\mathds{E}(K_n \mid H, \tilde{p})\} = \mathds{E}\left[\sum_{h=1}^H \{1 - (1 - W_h)^n\} \mid H\right].
\end{equation*}
where the quantity $1 - (1 - W_h)^n$ is the probability of observing the $h$th species at least once in a sample of size $n$ \citep{Smith1977} given the vector of probabilities $W_1,\dots,W_H$. Hence:
\begin{equation*}
\mathds{E}(K_n \mid H) = \mathds{E}\left[\sum_{h=1}^H \{1 - (1 - W_h)^n\} \mid H \right] = H - \mathds{E}\left\{\sum_{h=1}^H(1 - W_h)^n \mid H \right\} =H - H \mathds{E}\left\{(1 - W_1)^n \mid H \right\},
\end{equation*}
where the last step follows because  $W_1,\dots,W_H$ are iid beta random variables. In fact $(1 - W_1) \sim \text{Beta}(H |\sigma| - |\sigma|, |\sigma|)$ and from well-known properties of the beta distribution the result follows
\begin{equation*}
\mathds{E}(K_n \mid H) = H - H \frac{(H |\sigma| - |\sigma|)_n}{(H|\sigma|)_n}.
\end{equation*}
This proof technique can be extended to the posteriori expectation $\mathds{E}(K_m^{(n)} \mid H, X_1, \dots, X_n)$. In fact,  the posterior distribution of $W_1,\dots,W_H$ is \emph{conjugate}, that is
\begin{equation*}
(W_1,\dots,W_H \mid H, X_1,\dots,X_n) \sim \text{Dir}(|\sigma| + \bar{n}_1, \dots, |\sigma| + \bar{n}_H),
\end{equation*}
where $\bar{n}_1,\dots, \bar{n}_H$ are the labeled frequencies of the species $Z_1,\dots,Z_H$. Note that we could have $\bar{n}_h = 0$, and the number of non-zero entries, that is, the number of observed species, is $K_n = k$. Without loss of generality, we assume the non-zero abundances correspond to the first $k$ values, so that $n_j = \bar{n}_j$ for $j=1,\dots,k$ and $\bar{n}_j = 0$ for $j={k+1},\dots,H$. Reasoning as before, we obtain
\begin{equation*}
\begin{aligned}
\mathds{E}(K_m^{(n)} \mid H, X_1,\dots,X_n) &= \mathds{E}\{\mathds{E}(K_m^{(n)} \mid H, \tilde{p}, X_1,\dots,X_n)\} \\
&= \mathds{E}\left[\sum_{j=k+1}^H \{1 - (1 - W_j)^m\} \mid H, X_1,\dots,X_n\right]
\end{aligned}
\end{equation*}
where each $1 - (1 - W_j)^m$ for $j=k+1,\dots,H$ is the probability of sampling the $j$th unobserved species at least once in an additional sample of size $m$, given the vector of probabilities $W_1,\dots,W_H$ and the data $X_1,\dots, X_n$. A posteriori, the random variables $W_{k+1},\dots,W_H$ are still iid beta distributed and in particular $(1 - W_j \mid X_1,\dots,X_n) \sim \text{Beta}(n + H|\sigma|, |\sigma|)$ leading to
\begin{equation*}\begin{aligned}
\mathds{E}(K_m^{(n)} \mid H, X_1,\dots,X_n) &= H -k - (H -k)\mathds{E}\left\{(1 - W_{k+1})^m \mid H \right\} \\
&= H - k - (H-k)\frac{(n + H|\sigma| - |\sigma|)_m}{(n + H|\sigma|)_m}.
\end{aligned}
\end{equation*}
Interestingly, this corresponds to the estimator of a Pitman--Yor process with a negative discount parameter; see equation (6) in \citet{Favaro2009}.  We also note that this result could be alternatively obtained, with some effort, specializing the general theorems of \citet{Lijoi2007}. A further proof strategy based on combinatorial calculus is discussed in Appendix~A.1 and A.2 in \citet{Favaro2009}.

\subsection{Proof of Theorem~\ref{posterior_diversity}}

First of all, the quantity $ K_n = k $ is a sufficient statistic for the diversity by direct inspection of the likelihood function, that is, the \textsc{eppf}. In fact, the diversity $S_\sigma$ only appears in the terms $ V_{n,k}$ which solely depend on $n$ and $k$ and not the abundances. From Proposition~\ref{diversity}, we know that  
\begin{equation*}
\frac{K_n}{c_\sigma(n)} \overset{\text{a.s.}}{\longrightarrow} S_\sigma, \qquad n \to \infty,
\end{equation*}
that is, $\mathds{P}(\lim_{m\rightarrow \infty} K_m/c_\sigma(m) = S_\sigma) = 1$. Let $A = \left\{ \lim_{m\rightarrow \infty} K_m/c_\sigma(m) = S_\sigma \right\} $, and $ B = \{K_n = k\} $ and note that $\mathds{P}(B) > 0$ and $\mathds{P}(A) = 1$, therefore $\mathds{P}(A \cap B) = \mathds{P}(B)$. The proof follows from a very simple argument:
\begin{equation*}
\mathds{P}(A \mid B) = \frac{\mathds{P}(A \cap B)}{\mathds{P}(B)} = \frac{\mathds{P}(B)}{\mathds{P}(B)} = 1.
\end{equation*}
In other words, we have shown that $(K_m/c_\sigma(m) \mid K_n = k) \overset{\text{a.s.}}{\longrightarrow} S_\sigma$. By a continuity argument, the sequences 
\begin{equation*}
\left(\frac{K_{n+m}}{c_\sigma(m)} \mid K_n = k\right) \quad \text{and} \quad \left(\frac{K_{n+m}}{c_\sigma(n+m)} \mid K_n = k\right)
\end{equation*}
also converge to $ S_\sigma $ almost surely as $m \to \infty$. We now clarify that the distribution of the random variable $S_\sigma$ is indeed the posterior law of the parameters $H$, $\alpha$, and $\gamma$. Let us first consider the three building blocks of Gibbs-type priors, namely when $ S_\sigma $ is a positive constant. The above result implies the convergence of the Laplace functionals, that is, for any $ \lambda > 0 $,
\begin{equation*}
\mathds{E}\left(e^{-\lambda K_{n+m}/c_\sigma(m)} \mid K_n = k \right) \to e^{-\lambda S_\sigma}.
\end{equation*}
In a general Gibbs-type prior, $\left(K_{n+m}/c_\sigma(m) \mid K_n = k\right) $ converges to a random variable whose distribution is the posterior law of the diversity. In fact, as an application of the tower rule,
\begin{equation*}
\mathds{E}\left(e^{-\lambda K_{n+m}/c_\sigma(m)} \mid K_n = k \right) = \mathds{E}\left\{\mathds{E}\left(e^{-\lambda K_{n+m}/c_\sigma(m)} \mid K_n = k, S_\sigma \right)\right\} \to \mathds{E}(e^{-\lambda S_\sigma} \mid K_n = k),
\end{equation*}
which concludes the proof.

\subsubsection{Alternative proof of Theorem~\ref{posterior_diversity} for $\sigma = 0$}

We describe an alternative proof of Theorem~\ref{posterior_diversity}, expressed in terms of weak convergence, which applies when $\sigma = 0$. This proof technique is more direct, relying on calculus and combinatorics. In contrast, the previous proof is more abstract and depends on Proposition~\ref{diversity}.

Let us consider the Dirichlet process case with parameter $\alpha$. We begin by studying the a priori asymptotic behavior of $K_n / \log(n)$, a result that was established by \citet{Korwar1973}. By definition, the Laplace transform of this scaled random variable is
\begin{equation*}
\begin{aligned}
\mathds{E}\left(e^{-\lambda K_n / \log{n}}\right) &= \sum_{j=1}^n e^{ -\lambda j / \log{n}}  \frac{\alpha^j}{(\alpha)_n} |s(n, j)| = \frac{1}{(\alpha)_n} \sum_{j=1}^n \left(\alpha e^{- \lambda  / \log{n}}\right)^j   |s(n, j)| =   \\
&= \frac{(\alpha  e^{ -\lambda  / \log{n}})_n}{(\alpha)_n} = \frac{\Gamma(\alpha  e^{ -\lambda  / \log{n}} + n)}{\Gamma(\alpha  e^{ -\lambda  / \log{n}})} \frac{\Gamma(\alpha)}{\Gamma(\alpha + n)}.
\end{aligned}
\end{equation*}
The third equality follows by definition of signless Stirling numbers of the first kind. Now, let $g(n) = \alpha e^{-\lambda / \log(n)}$ and note that $\lim_{n\to\infty} g(n) = \alpha$. Thus
\begin{equation*}
\begin{aligned}
\lim_{n\to\infty}\mathds{E}\left(e^{-\lambda K_n / \log{n}}\right) &= \lim_{n\to\infty}\frac{\Gamma(g(n)+ n)}{\cancel{\Gamma(g(n))}} \frac{\cancel{\Gamma(\alpha)}}{\Gamma(\alpha + n)} = \lim_{n\to\infty}\frac{\Gamma(g(n) + n)}{\Gamma(\alpha + n)} \\
&= \lim_{n\to\infty} \frac{\sqrt{2\pi} e^{-g(n) - n}(g(n) + n)^{g(n) +n - 1/2}}{\sqrt{2\pi} e^{-\alpha - n}(\alpha + n)^{\alpha +n - 1/2}} \\
&=\lim_{n\to\infty} \frac{(g(n) + n)^{g(n) +n - 1/2}}{(\alpha + n)^{\alpha +n - 1/2}} \\
&=\lim_{n\to\infty} \left(\frac{(g(n) + n)^{g(n) +n - 1/2}}{(g(n) + n)^{\alpha +n - 1/2}}\right) \left(\frac{(g(n) + n)^{\alpha +n - 1/2}}{(\alpha + n)^{\alpha +n - 1/2}}\right) \\
&=\lim_{n\to\infty} (g(n) + n)^{g(n) - \alpha} \left(\frac{g(n) + n}{\alpha + n}\right)^{\alpha +n - 1/2}. 
\end{aligned}
\end{equation*}
The second step follows by applying Stirling's asymptotic formula for the Gamma function. Moreover, note that $\lim_{n\to\infty} (e^{-\lambda/\log{n}} - 1) / (-\lambda/\log{n}) = 1$ therefore
\begin{equation*}
\begin{aligned}
\lim_{n\to\infty} (g(n) + n)^{g(n) - \alpha} &= \lim_{n\to\infty} (\alpha e^{-\lambda/\log{n}} +n )^{\alpha(e^{-\lambda/\log{n}} - 1)} \\
&= \lim_{n\to\infty} (- \alpha \lambda/\log{n} +n )^{-\alpha\lambda/\log{n}} = e^{-\alpha\lambda},\\
\end{aligned}
\end{equation*}
whereas
\begin{equation*}
\begin{aligned}
\lim_{n\to\infty} \left(\frac{g(n) + n}{\alpha + n}\right)^{\alpha +n - 1/2} &= \lim_{n\to\infty} \frac{(1 + g(n)/n)^{\alpha +n - 1/2}}{(1 + \alpha/n)^{\alpha +n - 1/2}}
= \lim_{n\to\infty} e^{-\alpha}(1 + g(n)/n)^{\alpha +n - 1/2}\\
&= e^{-\alpha} e^\alpha = 1.
\end{aligned}
\end{equation*}
Summing up, we have shown that $\lim_{n\to\infty}\mathds{E}\left(e^{-\lambda K_n / \log{n}}\right) = e^{-\alpha \lambda}$, which means that $K_n / \log{n} \overset{\textup{d}}{\longrightarrow} \alpha$. The proof for the conditional case law follows from similar steps. Note that $(K_m^{(n)}/c_\sigma(m) \mid K_n = k)$ has the same asymptotic behavior as $(K_{n+m}/c_\sigma(m) \mid K_n = k)$, since $K_{n + m} = k + K_m^{(n)}$. Moreover, its Laplace transform is
\begin{equation*}
\begin{aligned}
\mathds{E}\left(e^{-\lambda K_m^{(n)} / \log{m}}\mid K_n = k\right) &= \sum_{j=1}^m \left(e^{ -\lambda / \log{m}}\right)^j \alpha^j \frac{(\alpha)_n}{(\alpha)_{n + m}} \sum_{\ell = j}^m\binom{m}{\ell}|s(\ell, j)| (n)_{m - \ell} \\
&=\frac{(\alpha)_n}{(\alpha)_{n + m}} \sum_{j=1}^m \left(\alpha e^{ -\lambda / \log{m}}\right)^j |s(m , j; n)| \\
&=\frac{(\alpha)_n}{(\alpha)_{n + m}} \left(\alpha e^{ -\lambda / \log{m}} + n\right)_m  
\end{aligned}
\end{equation*}
The first step follows from the combinatorial identity established in \citet{Lijoi2007}. In the second step, we recognized that $|s(m , j; n)| = \sum_{\ell = j}^m\binom{m}{\ell}|s(\ell, j)| (n)_{m - \ell}$ is an alternative representation of the signless non-central Stirling numbers of the first kind; see equation~(8.60) in \citet{Charalambides2002}. The third simplification follows by definition of $|s(m , j; n)|$. Taking the limit, we obtain
\begin{equation*}
\begin{aligned}
\lim_{m\to\infty} \mathds{E}\left(e^{-\lambda K_m^{(n)} / \log{m}}\mid K_n = k\right) &= \lim_{m\to\infty} \frac{\Gamma(\alpha +n)}{\Gamma(\alpha)}\frac{\Gamma(\alpha)}{\Gamma(\alpha + n + m)} \frac{\Gamma(\alpha e^{ -\lambda / \log{m}} + n)}{\Gamma(\alpha e^{ -\lambda / \log{m}} + n)} \\
&= \lim_{m\to\infty} \frac{\Gamma(\alpha e^{ -\lambda / \log{m}} + n)}{\Gamma(\alpha + n + m)} \\
&= e^{-\alpha \lambda}.
\end{aligned}
\end{equation*}
The last step follows from analogous calculations done for the a priori case. This concludes the proof, as the case of general Gibbs-type prior with random $\alpha$ and $\sigma = 0$ follows as an application of the tower rule, as before.

\subsection{Aldous-Pitman data augmentation}

In this Section we provide further details the discussion about the data augmentation for the Aldous-Pitman model described in Section~\ref{sec:ap}. Let us begin by recalling the main result, that is the augmented likelihood
\begin{equation*}
\Pi(n_1,\dots,n_k, u \mid \gamma) = \frac{2^{n - k/2 - 1/2}}{{\Gamma(2n - k - 1)}} \gamma^{k-1} u^{2n - k -2} e^{-u^2/2 - \sqrt{2}\gamma u} \prod_{j=1}^k(1/2)_{n_j-1}.
\end{equation*}
Integrating with respect to $u$ gives the \textsc{eppf} of an Aldous-Pitman model with weights in equation~\eqref{eq:ap}, because 
\begin{equation*}
h_{k+1-2n}(\sqrt{2}\gamma) = \frac{1}{\Gamma(2n - k - 1)}\int_0^\infty e^{-u^2/2 - \sqrt{2}\gamma} u^{2n -k - 2}\mathrm{d}u.
\end{equation*}
From the augmented representation we immediately obtain, through direct inspection of the joint law, the full conditional distributions for $\gamma$ and $U_{n,k}$. In particular:
\begin{equation*}
f_{U_{n,k}}(u \mid -) = \frac{e^{-u^2/2 - \sqrt{2}\gamma} u^{2n -k - 2}}{\int_0^\infty e^{-u^2/2 - \sqrt{2}\gamma} u^{2n -k - 2} \mathrm{d}u}.
\end{equation*}
Moreover, let us recall the predictive distribution is
\begin{equation*}
\mathds{P}(X_{n+1} \in \cdot \mid \gamma, X_1,\dots,X_n) = \sqrt{2}\gamma \frac{h_{k-2n}(\sqrt{2}\gamma)}{h_{k+1-2n}(\sqrt{2}\gamma)}P(\cdot) + 2\frac{h_{k-1-2n}(\sqrt{2}\gamma)}{h_{k + 1 - 2n}(\sqrt{2}\gamma)}\sum_{j=1}^k(n_j - 1/2)\delta_{X^*_j}(\cdot).
\end{equation*}
Substituting $h_{k+1-2n}(\sqrt{2}\gamma)$ and $h_{k-2n}(\sqrt{2}\gamma)$ with their integral representation, we obtain
\begin{equation*}\begin{aligned}
\sqrt{2}\gamma \frac{h_{k-2n}(\sqrt{2}\gamma)}{h_{k+1-2n}(\sqrt{2}\gamma)} &= \sqrt{2}\gamma \frac{\Gamma(2n - k - 1)}{\Gamma(2n - k)}\frac{\int_0^\infty u^{2n-k-1} e^{-u^2/2 - \sqrt{2}\gamma u}\mathrm{d}u}{\int_0^\infty u^{2n-k-2} e^{-u^2/2 - \sqrt{2}\gamma u}\mathrm{d}u} \\
&= \frac{\sqrt{2}\gamma}{2n - k - 1}\int_0^\infty u f_{U_{n,k}}(u \mid -) \mathrm{d}u \\
&= \frac{\sqrt{2}\gamma}{2n - k - 1} \mathds{E}(U_{n,k}).
\end{aligned}
\end{equation*}
and 
\begin{equation*}\begin{aligned}
2\frac{h_{k-1-2n}(\sqrt{2}\gamma)}{h_{k+1-2n}(\sqrt{2}\gamma)} &= 2 \frac{\Gamma(2n - k - 1)}{\Gamma(2n - k + 1)}\frac{\int_0^\infty u^{2n-k} e^{-u^2/2 - \sqrt{2}\gamma u}\mathrm{d}u}{\int_0^\infty u^{2n-k-2} e^{-u^2/2 - \sqrt{2}\gamma u}\mathrm{d}u} \\
&= \frac{2}{(2n -k)(2n -k -1)}\int_0^\infty u^2 f_{U_{n,k}}(u \mid -) \mathrm{d}u \\
&=\frac{2}{(2n -k)(2n -k -1)}\mathds{E}(U^2_{n,k}).
\end{aligned}
\end{equation*}
Note that this predictive scheme still does not provide a manageable sampling algorithm for $X_{n+1} \mid X_1, \dots, X_n$ because the expectations $\mathds{E}(U_{n,k})$ and $\mathds{E}(U^2_{n,k})$ are not easily available. However, one can resort to Algorithm~\ref{algo1}, which is based on the following additional data augmentation. Let us define a binary random variable $D_{n+1}$ such that $D_{n+1} = 1$ iff $X_{n+1} = \text{``new''}$ and $0$ otherwise. Moreover, let $\tilde{U}_{n,k}$ be a positive random variable such that the joint law of $D_{n+1}$ and $\tilde{U}_{n,k}$ is:
\begin{equation*}
\mathds{P}(D_{n+1} = 1, \tilde{U}_{n,k} \in \mathrm{d}u) = \frac{\sqrt{2}\gamma}{2n - k - 1}u f_{U_{n,k}}(u \mid -) \mathrm{d}u,
\end{equation*}
and 
\begin{equation*}
\mathds{P}(D_{n+1} = 0, \tilde{U}_{n,k} \in \mathrm{d}u) = \frac{u^2}{2n -k -1} f_{U_{n,k}}(u \mid -).
\end{equation*}
Then, the sampling strategy is as follows: sample $\tilde{U}_{n,k}$ from its marginal distribution and then $D_{n+1} \mid \tilde{U}_{n,k}$. Both distributions are easy to sample from and their laws are described in Algorithm~\ref{algo1}. Once it has been determined whether the observation $X_{n+1}$ is either new or old, the remaining part of the sampling algorithm is trivial.% - Predictive distribution, derivation
% - Algorithm

\begin{algorithm}[tb]
\caption{Sampling procedure for $(X_{n+1} \mid X_1,\dots,X_n)$ for an Aldous-Pitman model. \label{algo1}}
\begin{algorithmic}
\Require $X_1,\dots,X_n$.
\State 1. Sample a positive latent variable $\tilde{U}_{n, k}$ from the log-concave density $$
f_{\tilde{U}_{n,k}}(u) \propto \left(\frac{\gamma}{\sqrt{2}} \frac{u}{(n - k/2 - 1/2)} + \frac{u^2}{(2n - k - 1)}\right)u^{2n - k -2} e^{-(u/\sqrt{2}  + \gamma) ^2}.
$$
\State 2. Sample a binary random variable $D_{n + 1}$ representing if $X_{n+1}$ is new or not according to
$$
\mathds{P}(D_{n+1} = 1 \mid \tilde{U}_{n, k} = u) \propto \frac{\gamma}{\sqrt{2}} \frac{u}{(n - k/2 - 1/2)}, \quad \mathds{P}(D_{n+1} = 0 \mid \tilde{U}_{n, k} = u) \propto \frac{u^2}{(2n - k - 1)}.
$$
\State 3. Sample $X_{n+1}$ given the binary variable $D_{n+1}$:
\If{$D_{n, k} = 1$ (i.e. if $X_{n+1} = \text{``new''})$} 
    \State 3.a Sample a new value $X_{n + 1}$ from the baseline $P$;
\ElsIf{$D_{n, k} = 0$ (i.e. if $X_{n+1} = \text{``old''})$}
    \State 3.b Sample $X_{n + 1}$ from the discrete distribution $\mathds{P}(X_{n+1} = X^*_j \mid D_{n+1} = 0) \propto (n_j - 1/2)$.
\EndIf

\State \Return the sampled value $X_{n+1}$.
\end{algorithmic}
\end{algorithm}

\subsection{Proof of Proposition~\ref{prop:taxon_urn}}

The proof of this proposition follows by repeatedly applying the argument discussed in \citet{Rigon2025} for the enriched Pitman-Yor process, corresponding to the case where $L = 2$ and the involved random measures are either Pitman--Yor processes or Dirichlet processes. 

\subsection{Details of the Gibbs sampling algorithm of Section~\ref{sec:app2}}

The augmented and coarsened likelihood of the model Section~\ref{sec:app2} is the following

\begin{equation*}
\begin{aligned}
\mathscr{L}(\bm{X}_1,\dots,\bm{X}_n \mid \bm{S}, \bm{U}) \propto \underbrace{\frac{\alpha_1^{k_1}}{(\alpha_1)_n}}_\textup{layer 1: families}   &\times \overbrace{\prod_{j=1}^{k_1} \frac{\alpha(X^*_{j, 1})^{k(X^*_{j,1})}}{(\alpha(X^*_{j, 1}))_{n(X^*_{j,1})}}}^\textup{layer 2: genera within family} \\
&\times \underbrace{\prod_{j=1}^{k_2} \left[\gamma(X^*_{j,2})^{k(X^*_{j,2})-1} u_j^{2n(X^*_{j,2}) - k(X^*_{j,2}) -2} e^{-u_j^2/2 - \sqrt{2}\gamma(X^*_{j,2}) u_j}\right]^{\rho}}_\textup{layer 3: species within genus},
\end{aligned}
\end{equation*}
where $\bm{U} = (u_1, \dots, u_{k_2})$ represents the collection of latent variables. Note that the likelihood function factorizes, allowing inference to be performed independently for each layer of the model. For layer 2, inference proceeds separately for each  $\alpha(X^*_{j, 1})$ as described in Section~\ref{sec:app1}, given that the priors for the diversities $\alpha(X^*_{j, 1})$ are independently distributed according to the Stirling-gamma distribution. On the other hand, a Gibbs-sampling algorithm is required for layer 3, alternating between these steps
\begin{enumerate}
\item For $j=1,\dots,k_2$ sample independently from the (coarsened) full conditionals of the diveristies $(\gamma(X^*_{j,2}) \mid -)\sim \text{Gamma}(a_\gamma + \rho\{k(X^*_{j,2}) - 1\}, b_\gamma + \rho \sqrt{2} u_j)$.

\item For $j=1,\dots,k_2$ sample independently from the (coarsened) full conditionals of $(U_j \mid -)$ whose densities are $f_{U_j}(u) \propto u^{\rho 2n(X^*_{j,2}) - k(X^*_{j,2}) -2} e^{-\rho u^2/2 - \rho  \sqrt{2}\gamma(X^*_{j,2}) u}$. Exact sampling from this distribution is feasible using the algorithm described in \citet{Sun2023}, or alternatively, through the ratio-of-uniform acceptance-rejection algorithm.

\item Sample the hyperparameters $(\log{a_\gamma}, \log{b_\gamma} \mid -)$ from their full conditional distribution. This step is equivalent to sampling the posterior distribution of the parameters under the assumption that the data are i.i.d. Gamma distributed and the log-prior follows a multivariate Gaussian distribution. A standard Metropolis step is employed, using a carefully tuned Gaussian proposal distribution to ensure good mixing.
\end{enumerate}

\section{Additional plots for the application in Section~\ref{sec:app1}}

Additional plot for the application discussed in Section~\ref{sec:app1}. Simulated values are based on $10^6$ Monte Carlo replicates using the Stirling-gamma sampling algorithm of \citet{Zito2024} combined with a Poisson approximation for $K_N$, as discussed in the main text.

\begin{figure}[tbp]
\centering
    \includegraphics[width=0.7\textwidth]{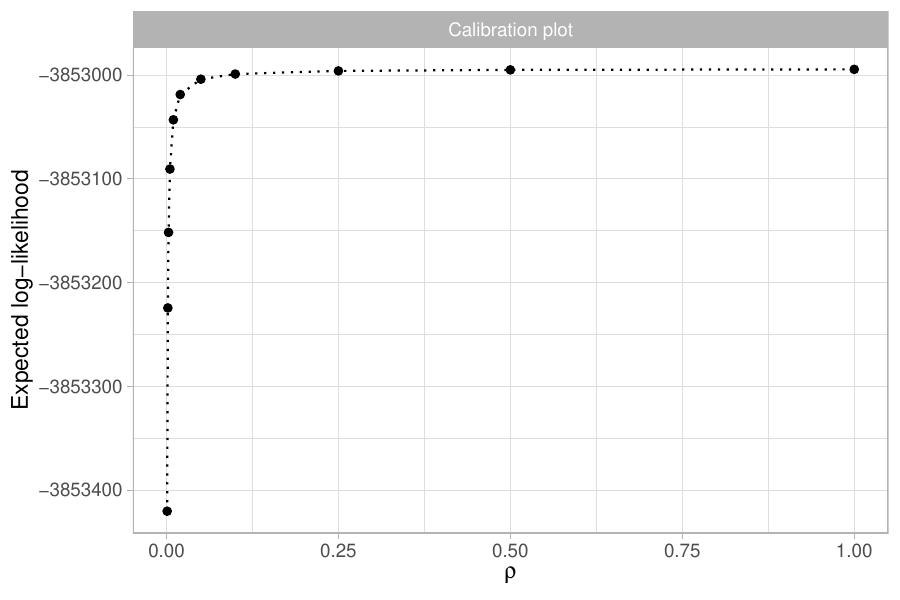}
    \caption{Calibration plot comparing the coarsening level $\rho$ and the expected log-likelihood $\mathds{E}\{k \log{\alpha} - \log{(\alpha)_n}\}$, where the expectation is taken with respect to the coarsened posterior of $\alpha$, for various level of $\rho$.}
    \label{fig:calibration_alpha}
\end{figure}

\begin{figure}[tbp]
\centering
    \includegraphics[width=0.7\textwidth]{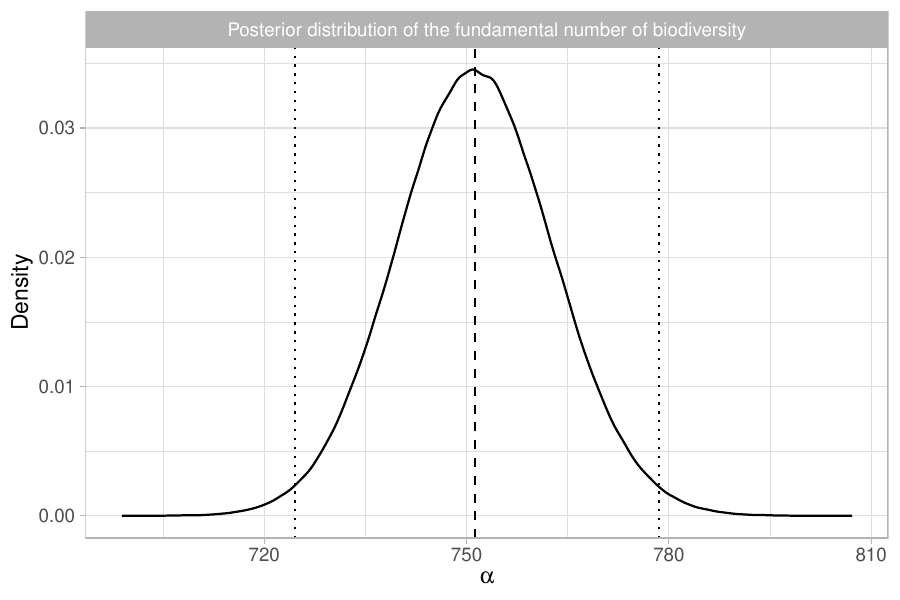}
    \caption{Posterior distribution ($\rho = 1$) of the fundamental biodiversity number $\alpha$, using the Amazonian tree dataset of \citet{TerSteege2013}. The dotted lines represent $98\%$ credible intervals. The dashed line is the posterior mean.}
    \label{fig:post_alpha_power}
\end{figure}

\begin{figure}[tbp]
\centering
    \includegraphics[width=0.7\textwidth]{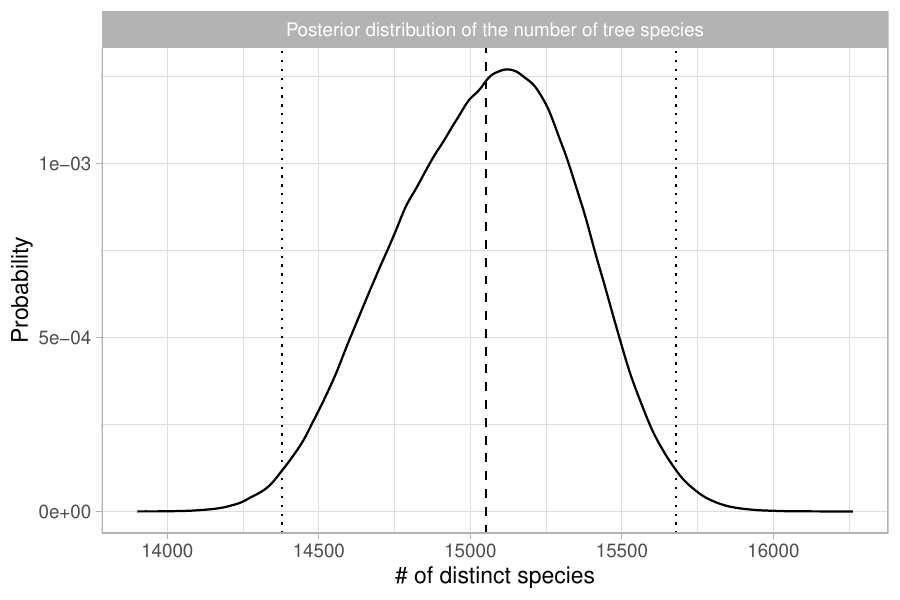}
    \caption{Posterior distribution ($\rho = 1$) of the total number of distinct tree species $K_N$, using the Amazonian tree dataset of \citet{TerSteege2013}. The dotted lines represent $98\%$ credible intervals. The dashed line is the posterior mean. }
    \label{fig:total_species}
\end{figure}

%\begin{figure}[tbp]
%\centering
%    \includegraphics[width=0.8\textwidth]{alpha_power.pdf}
%    \caption{asd}
%    \label{fig:total_species}
%\end{figure}

\begin{figure}[tbp]
\centering
    \includegraphics[width=0.7\textwidth]{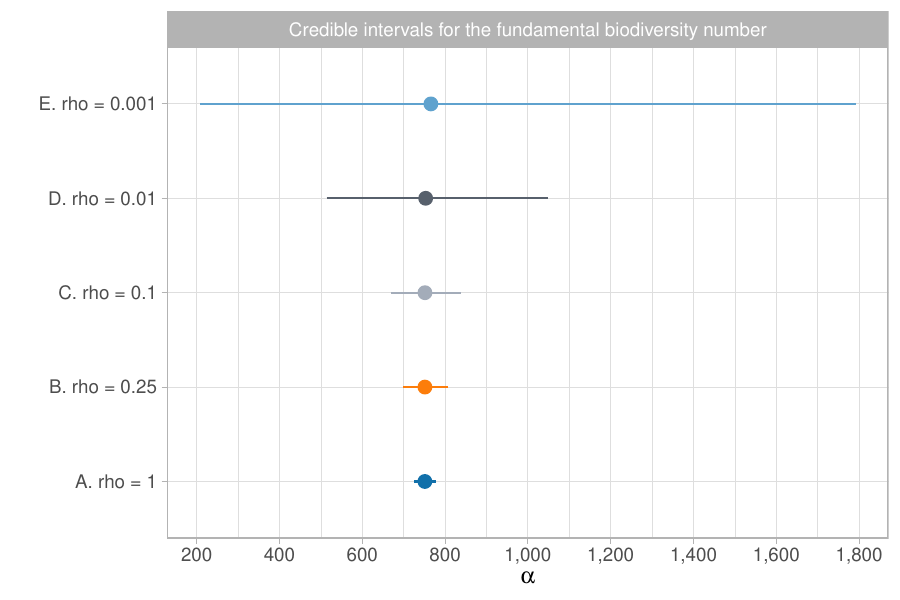}
    \caption{Credible intervals of the posterior distribution of the fundamental biodiversity number, using the Amazonian tree dataset of \citet{TerSteege2013}, for various choices of $\rho \in \{0.001, 0.01, 0.1, 0.25, 1\}$. }
    \label{fig:total_species}
\end{figure}

%\begin{figure}[tbp]
%\centering
%    \includegraphics[width=0.7\textwidth]{total_species_power.pdf}
%    \caption{asd}
%    \label{fig:total_species}
%\end{figure}

\begin{figure}[tbp]
\centering
    \includegraphics[width=0.7\textwidth]{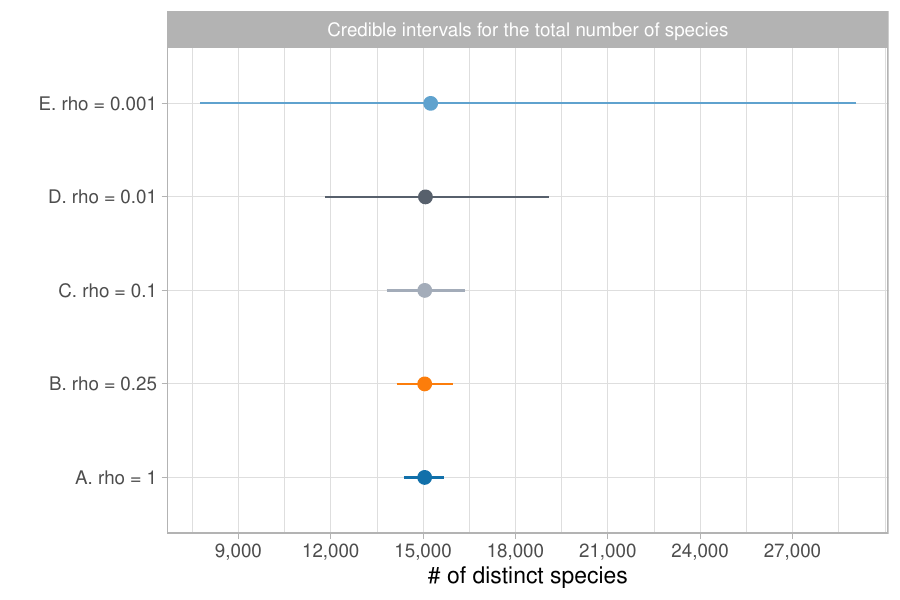}
    \caption{Credible intervals of the posterior distribution of the total number of distinct tree species $K_N$, using the Amazonian tree dataset of \citet{TerSteege2013}, for various choices of $\rho \in \{0.001, 0.01, 0.1, 0.25, 1\}$. }
    \label{fig:total_species}
\end{figure}

\end{document}